\documentclass[12pt,preprint]{aastex}

\defcitealias{stern12}{Paper~I}

\usepackage{natbib}
\usepackage{amsmath}

\title{Mid-Infrared Selection of Active Galactic Nuclei with the
  Wide-Field Infrared Survey Explorer. II. Properties of WISE-Selected
  Active Galactic Nuclei in the NDWFS Bo\"otes Field}

\author{R.J.~Assef\altaffilmark{1,2}, 
  D.~Stern\altaffilmark{1}, 
  C.S.~Kochanek\altaffilmark{3,4},
  A.W.~Blain\altaffilmark{5},
  M.~Brodwin\altaffilmark{6},
  M.J.I.~Brown\altaffilmark{7},
  E.~Donoso\altaffilmark{8,9},
  P.R.M.~Eisenhardt\altaffilmark{1},
  B.T.~Jannuzi\altaffilmark{10},
  T.H.~Jarrett\altaffilmark{8,11},
  S.A.~Stanford\altaffilmark{12,13},
  C.-W.~Tsai\altaffilmark{1,2},
  J.~Wu\altaffilmark{14},
  L.~Yan\altaffilmark{8}
}

\altaffiltext{1}{Jet Propulsion Laboratory, California Institute of
  Technology, 4800 Oak Grove Drive, Mail Stop 169-236, Pasadena, CA
  91109, USA. E-mail:~{\tt{roberto.j.assef@jpl.nasa.gov}}}
    
\altaffiltext{2}{NASA Postdoctoral Program Fellow}

\altaffiltext{3}{Department of Astronomy, The Ohio State University,
  140 W.\ 18th Ave., Columbus, OH 43210, USA}

\altaffiltext{4}{The Center for Cosmology and Astroparticle Physics,
  The Ohio State University, 191 West Woodruff Avenue, Columbus, OH
  43210, USA}

\altaffiltext{5}{Physics \& Astronomy, University of Leicester, 1
  University Road, Leicester, LE1 7RH, UK}

\altaffiltext{6}{Department of Physics, University of Missouri, 5110
  Rockhill Road, Kansas City, MO 64110, USA}

\altaffiltext{7}{School of Physics, Monash University, Clayton 3800,
  Victoria, Australia}

\altaffiltext{8}{Infrared Processing and Analysis Center, California
  Institute of Technology, Pasadena, CA 91125, USA}

\altaffiltext{9}{Instituto de Ciencias Astronomicas, de la Tierra, y
  del Espacio (ICATE), 5400 San Juan, Argentina}

\altaffiltext{10}{Steward Observatory, University of Arizona, Tucson,
  AZ 85721, USA}

\altaffiltext{11}{Astronomy Department, University of Cape Town,
  Rondebosch, South Africa}

\altaffiltext{12}{Department of Physics, University of California, One
  Shields Avenue, Davis, CA 95616, USA}

\altaffiltext{13}{Institute of Geophysics and Planetary Physics,
  Lawrence Livermore National Laboratory, Livermore, CA 94550, USA}

\altaffiltext{14}{UCLA Astronomy, PO Box 951547, Los Angeles, CA
  90095-1547, USA}

\begin{document}

\begin{abstract}
\citet{stern12} presented a study of WISE selection of AGN in the 2
deg$^2$ COSMOS field, finding that a simple criterion W1--W2$\geq$0.8
provides a highly reliable and complete AGN sample for W2$<$15.05,
where the W1 and W2 passbands are centered at 3.4$\mu$m and 4.6$\mu$m,
respectively. Here we extend this study using the larger 9 deg$^2$
NOAO Deep Wide-Field Survey Bo\"otes field which also has considerably
deeper WISE observations than the COSMOS field, and find that this
simple color-cut significantly loses reliability at fainter fluxes. We
define a modified selection criterion combining the W1$-$W2 color and
the W2 magnitude to provide highly reliable or highly complete AGN
samples for fainter WISE sources. In particular, we define a
color-magnitude cut that finds 130$\pm$4~deg$^{-2}$ AGN candidates for
W2$<$17.11 with 90\% reliability. Using the extensive UV through
mid-IR broad-band photometry available in this field, we study the
spectral energy distributions of WISE AGN candidates. We find that, as
expected, the WISE AGN selection can identify highly obscured AGN, but
that it is biased towards objects where the AGN dominates the
bolometric luminosity output. We study the distribution of reddening
in the AGN sample and discuss a formalism to account for sample
incompleteness based on the step-wise maximum-likelihood method of
\citet{efstathiou88}. The resulting dust obscuration distributions
depend strongly on AGN luminosity, consistent with the trend expected
for a \citet{simpson05} receding torus. At $L_{\rm AGN} \sim3\times
10^{44}~\rm erg~\rm s^{-1}$, 29$\pm$7\% of AGN are observed as Type 1,
while at $\sim 4\times 10^{45}~\rm erg~\rm s^{-1}$ the fraction is
64$\pm$13\%. The distribution of obscuration values suggests that dust
in the torus is present as both a diffuse medium and in optically
thick clouds.
\end{abstract}


\section{Introduction}

Active galactic nuclei (AGN) have been proposed to play an
important role in several aspects of galaxy evolution, such as
quenching star-formation in their host galaxies by heating and/or
mechanically pushing their gas reservoirs into the intergalactic
medium \citep[see, e.g.,][]{hopkins05}, preventing cooling flows at
the center of galaxy clusters \citep[see, e.g.,][]{croton06}, and
possibly by contributing significantly to the reionization of the
Universe at high redshift \citep{glikman10,glikman11}. Efficiently
identifying AGN in all states of accretion and obscuration and
accurately understanding their properties and structure is a key step
to understand how galaxies evolve with cosmic time.

AGN are among the most luminous objects in the Universe. Most of the
radiated energy is thermally generated by the accretion disk
surrounding the central super-massive black-hole (SMBH) within scales
of $\sim 1~\rm AU$, with a spectrum that is well approximated by a
declining power-law shortwards of soft X-ray wavelengths
\citep[e.g.,][]{shakura73}. A considerable fraction of this luminosity
is absorbed by dust surrounding the SMBH on scales of $\sim 1~\rm pc$,
which is heated up to temperatures reaching its sublimation limit
($\sim 1500~\rm K$) and re-radiates it in the IR. The dust
distribution is typically thought to have a quasi-toroidal shape
\citep[see, e.g.,][]{urry95}, with a scale height and opening angle
that depends on the luminosity of the accretion disk \citep[see,
  e.g.,][]{simpson05}. The dust emission dominates the mid-IR emission
in an AGN and rises as a power-law towards longer wavelengths,
providing AGN with their characteristic red mid-IR colors
\citep[e.g.][]{elvis94,stern05,richards06,assef10} that allows their
identification even when the accretion disk emission is blocked by the
dust torus, as it differs strongly from the Rayleigh-Jeans emission of
the stellar population that dominates the mid-IR spectrum of inactive
galaxies.

The Wide-field Infrared Survey Explorer \citep[WISE;][]{wright10} is a
NASA satellite with a 40cm aperture that imaged the whole sky in four
mid-IR bands, centered at 3.4, 4.6, 12 and 22$\mu$m. We refer to these
bands as W1, W2, W3 and W4, respectively. The fully cryogenic WISE
science mission started in January 2010 and ended in August of the
same year; all of this data has been publicly available since March
2012. While not one of its main goals, WISE is well suited to studying
AGN as its bands are sensitive to their characteristic warm dust
emission and are little affected by the obscuration expected from
either the dust in the torus or in the interstellar medium of the host
galaxy. In a companion work, \citet[][\citetalias{stern12}]{stern12}
investigated the power of WISE to identify AGN based solely on the
W1--W2 mid-IR color by comparing to known AGN in the COSMOS field. The
selection also necessarily entails a flux cut, which is relatively
shallow given the low ecliptic latitude of the COSMOS field and thus
lower WISE coverage depth. Using a set of AGN in the field selected
from the {\it{Spitzer}} Infrared Array Camera \citep[IRAC;][]{fazio04}
colors according to the criteria developed by \citet{stern05} as a
control sample, we determined that at the depth of the COSMOS field a
very simple selection criterion of W1$-$W2$\geq$0.8 and W2$<$15.05
produced an AGN sample with a contamination of only 5\% and recovered
nearly 80\% of the IRAC-selected AGN in the field to that WISE
depth. The $61.9\pm5.4$ deg$^{-2}$ space density of these AGN is about
three times higher than that of similar bolometric luminosity Type 1
AGN found at optical wavelengths by the Sloan Digital Sky Survey
\citep[SDSS;][]{richards02}. The color of this criterion is similar to
that proposed earlier by \citet[][W1--W2$>$0.85]{assef10} using mock
WISE data constructed from SED models calibrated in this wavelength
range by {\it{Spitzer}} data. \citet{assef10} showed that WISE and
SDSS are sensitive to AGN of the same bolometric luminosities for
$z<4$, implying that the increased census is due to the sensitivity of
WISE to those objects obscured by dust. In fact, \citetalias{stern12}
showed that the distribution of the X-ray hardness ratios of the
WISE-selected AGN are, as expected, consistent with a considerable
number of dust-obscured AGN.

Because of the polar orbit and near continuous observing strategy of
WISE, the depth of a field depends strongly on ecliptic latitude. The
2 deg$^2$ COSMOS field, close to the ecliptic, is representative of
the shallowest WISE fields. In this work we extend the work presented
in \citetalias{stern12} to the much larger, 9 deg$^2$ NOAO Deep
Wide-Field Survey \citep[NDWFS;][]{jannuzi99} Bo\"otes field, which
has considerably deeper WISE observations due to its higher ecliptic
latitude. In \S\ref{sec:data} we describe the data sets and spectral
energy distribution (SED) fitting models which we use to extend the
AGN selection criteria of \citetalias{stern12} to fainter WISE fluxes
in \S\ref{sec:wise_agn_sel}. In \S\ref{sec:wise_agn_sed_properties} we
study the broad-band SEDs of the WISE AGN candidates, and assess the
accuracy with which we can estimate photometric redshifts for
them. Finally, in \S\ref{sec:dust_properties}, we study the
distribution of the obscuring dust in AGN and present a method to
correct for sample incompleteness due to reddening. Throughout this
work we assume a flat $\Lambda$CDM cosmology with $H_0 = 73~\rm km~\rm
s^{-1}$, $\Omega_M=0.3$ and $\Omega_{\Lambda}=0.7$. We refer to all
magnitudes in their native photometric system, i.e., AB for
{\it{ugriz}}, FUV and NUV, and A0 (Vega) for all other bands.

\section{Data and Modeling}\label{sec:data}

\subsection{The NOAO Deep Wide-Field Survey Bo\"otes Field}\label{ssec:ndwfs}

NDWFS is a deep imaging survey in $B_W$, $R$, $I$ and $K$ of two 9
deg$^2$ fields in the constellations of Cetus and Bo\"otes. We focus
here on the Bo\"otes field, for which follow-up deep imaging has been
obtained for a wide range of wavelengths. Bo\"otes also has deep and
extensive spectroscopy.

Follow-up imaging of the Bo\"otes field exists from the X-rays with
{\it{Chandra}} \citep[XBo\"otes;][]{murray05} to the radio from the
Faint Images of the Radio Sky at Twenty-centimeters
\citep[FIRST;][]{becker95} survey, the NRAO VLA Sky Survey
\citep[NVSS;][]{condon98}, the Westerbork Northern Sky Survey
\citep[WENSS;][]{rengelink97} and from \citet{devries02}. The whole
field was observed with 90 seconds of exposure per position in the
IRAC Shallow Survey \citep{eisenhardt04}. The {\it{Spitzer}} Deep,
Wide-Field Survey \citep[SDWFS;][]{ashby09} quadrupled this exposure,
reaching 5$\sigma$ depths of 19.3, 18.5, 16.3 and 15.6~mag for [3.6],
[4.5], [5.8] and [8.0], respectively. Additionally, we also use the
{\it{Galaxy Evolution Explorer}} \citep[{\it{GALEX}};][]{martin05}
Deep Imaging Survey (DIS) and All-sky Imaging Survey (AIS) FUV and NUV
observations of the field, the $z$-band data of \citet{cool07}, the
near-IR $J$, $H$ and $K_s$ observations of NEWFIRM \citep{gonzalez10}
and the MIPS 24$\mu$m observations of the MIPS AGN and Galaxy
Evolution Survey \citep[MAGES;][]{jannuzi10}. For our work, we use
6\arcsec\ aperture magnitudes, corrected for PSF losses, obtained from
PSF-matched images in all but the {\it{Spitzer}} bands.

The AGN and Galaxy Evolution Survey \citep[AGES;][]{kochanek12}
obtained deep optical spectra of approximately 25,000 sources in the
Bo\"otes field with Hectospec \citep{fabricant05} at the Multiple
Mirror Telescope (MMT). The survey is designed to be statistically
complete for several different samples limited to $I<20$ for galaxy
candidates and $I<22.5$ for AGN candidates. AGES is highly complete
for AGN candidates to $I<21.5$ \citep[see][for details on completeness
  and selection]{kochanek12}. The AGN candidates were targeted by
their X-ray, radio and mid-IR properties, but not by their optical
colors. This ensures that none of the optical selection biases
\citep[see, e.g.,][]{fan99} are propagated into the sample. We
complement the AGES spectroscopy with $\sim2000$ deeper optical
spectra from various sources, primarily from Keck
\citep[e.g.,][]{eisenhardt08}. Since these data do not have a uniform
selection function, they will only be of limited use in our analysis.

\subsection{WISE Observations}\label{ssec:data_wise}

The WISE mission observed the full sky in four mid-IR photometric
bands with FWHM of 6\arcsec\ in W1--3 and 12\arcsec\ in W4. We use the
WISE all-sky data release, which includes all observations obtained
during the fully cryogenic mission. WISE surveyed the sky in a polar
orbit with respect to the ecliptic, simultaneously obtaining images in
all four bands. Hence, the number of observations in a field increases
with its ecliptic latitude. While fields near the ecliptic were
typically observed 12 times, the number can grow to several hundreds
near the ecliptic poles \citep[e.g.,][]{jarrett11}. The median
coverage across the sky is approximately 15 frames per passband. In
particular, the COSMOS field was observed with a median coverage of 11
frames per passband, well below the median sky coverage. Detailed
accounts of the mission are presented by \citet{wright10} and in the
WISE all-sky data release explanatory
supplement\footnote{\url{http://wise2.ipac.caltech.edu/docs/release/allsky/expsup/}}.

The NDWFS Bo\"otes field is at an ecliptic latitude of 46 deg, and
hence WISE obtained an average coverage of 30 frames in each band,
reaching 10$\sigma$ depths in W1, W2, W3 and W4 of approximately
17.12~mag, 15.73~mag, 11.55~mag and 7.83~mag. For all sources, we use
fluxes obtained through profile fitting. We limit the sample to
$S/N>3$ in W1 and W2, or equivalently to W1$<$18.50 and W2$<$17.11. We
match to other sources in Bo\"otes by finding the closest IRAC [4.5]
source within 2\arcsec\ with the constraint that no WISE (IRAC) source
is matched to more than one IRAC (WISE) source. This results in a
sample of 111,720 matched sources. We note that the WISE magnitude
limit is applied after cross-matching with the IRAC sources.

Detailed comparison between WISE and {\it{Spitzer}} IRAC photometry
has shown that WISE profile-fitting fluxes in W1 and W2 are typically
underestimated for faint sources, and that the magnitude of the effect
increases with decreasing IRAC flux, reaching offsets of a few tenths
of a magnitude for the fainter sources (see section VI.3 of the WISE
all-sky data release explanatory supplement for details). No similar
effect is observed for W3 and W4. While there is no simple
prescription to mitigate it, this bias is unlikely to affect the
results of our SED fits, as it is only significant for faint sources
for which the deeper IRAC SDWFS magnitudes dominate the $\chi^2$ of
the fit in the mid-IR. Because of this, we do not attempt to
compensate for this WISE calibration issue in our study.

\subsection{Spectral Energy Distribution Modeling}\label{ssec:sed_fits}

We rely on SED modeling both to obtain physical insight into our AGN
candidates, and to obtain photometric redshift ($z_{\rm phot}$)
estimates for all objects without available spectroscopic
redshifts. To fit the SEDs we use the non-negative basis of
low-resolution, UV through mid-IR SED templates for AGN and galaxies
of \citet{assef10}. The basis consists of four empirically derived SED
templates, where every object is modeled as a non-negative combination
of the three galaxy SED templates (roughly corresponding to E, Sbc and
Im types) and the single AGN template. For the AGN template alone, we
allow reddening with a strength parametrized by $E(B-V)$. For
high-redshift sources, we model the intergalactic medium (IGM)
absorption following \citet{fan06} for Ly$\alpha$ and Ly$\beta$
absorption and \citet{stengler95} for Lyman limit systems. The
strength of the IGM absorption can also be fit beyond the standard
mean absorption law, although this extra degree of freedom often has a
negative impact on the accuracy of photometric redshifts. A weak prior
is used to keep $E(B-V)$ as small as possible with the secondary
effect that obscuration values may be slightly underestimated in some
cases. This prior is required to avoid an extremely reddened AGN
component being used to improve the fit to the longest wavelength
bands (primarily W3, W4 and MIPS 24$\mu$m) in an inactive galaxy
without affecting the SED at shorter wavelengths. Also, this prior can
lower the possible degeneracy between a red stellar spectrum and a
reddened AGN in $z\gtrsim 1$ inactive galaxies with little or no
rest-frame mid-IR constraints, although we note this is very unlikely
to happen in our W2 selected sample given the deep SDWFS IRAC [5.8]
and [8.0] observations. We refer the reader to \citet{assef10} for
further details on the $E(B-V)$ prior.

We follow the prescription detailed in \citet{assef10} to obtain
photometric redshifts and fit the SEDs. Since photometric redshifts
using 24$\mu$m photometry have lower accuracies when using these
templates \citep[see][for details]{assef10}, we derive photometric
redshifts for all objects in the sample using only the broad-band
photometry from FUV to W3. Adding MIPS 24$\mu$m and W4 photometry,
however, does not qualitatively alter our results. We assume the
standard mean IGM absorption and use a luminosity prior for the galaxy
components based on the Las Campanas Redshift Survey (LCRS) $r-$band
luminosity function \citep{lin96}. We discuss the precision of the
photometric redshifts in \S\ref{ssec:photoz_acc}.

After obtaining photometric redshift estimates, we re-fit the SEDs of
all objects now including the W4 and MIPS 24$\mu$m channels, and also
fit for the strength of the IGM absorption. Whenever possible, we use
spectroscopic redshifts ($z_{\rm spec}$). This approach ensures that
we get the best SED model possible for each object. Several authors
have determined that photometric redshifts for Type 1 AGN obtained
solely with broad-band filter photometry can be wildly inaccurate
\citep[see, e.g.,][]{rowan-robinson08,salvato09,assef10}; this is
discussed further in \S\ref{ssec:photoz_acc} in the context of our
study. However, our spectroscopic data is particularly deep and
complete for AGN (see \S\ref{ssec:ndwfs}), somewhat mitigating this
issue.

In order to reliably separate AGN from inactive galaxies, we use the
parameter
\begin{equation}\label{eq:ahat}
  \hat{a}\ \equiv\ \frac{L_{\rm AGN}}{L_{\rm host} + L_{\rm AGN}},
\end{equation}
\noindent where the luminosities correspond to the integrated specific
luminosities of the best-fit templates over the 0.1 to 30$\mu$m
wavelength range for the AGN template and 0.03 to 30 $\mu$m for the
host galaxy templates \citep[see][for details]{assef10}. The specific
luminosities are calculated after correcting the AGN component for the
best-fit value of the reddening. We refer to these as bolometric
luminosities for the rest of the paper. Note that \citet{assef10}
determined that $\hat{a}$ is insensitive to photometric redshift
uncertainties as long as enough data exists to constrain the fit, in
the sense that $\hat{a}$ can still be accurately determined for
objects where reliable photometric redshifts cannot be measured. That
is to say, while it is challenging to measure accurate photometric
redshifts for AGN, particularly for Type 1 AGN, we are able to
accurately disentangle the relative fractions of starlight and nuclear
emission even when the redshift estimate is significantly in error. A
general characterization of this accuracy beyond that in
\citet{assef10} is presented in Appendix A.

\section{WISE AGN Color Selection}\label{sec:wise_agn_sel}

In this section we study the completeness and reliability of WISE AGN
selection. First, we discuss the criterion of \citetalias{stern12}
applied to the deeper WISE data in the Bo\"otes field, while in
\S\ref{ssec:mag_dep_cut} we improve our method by also considering the
observed W2 magnitude of the sources. In \S\ref{ssec:lit_comp} we
compare our new method with others in the literature.

\subsection{Magnitude-Independent AGN Color Selection}\label{ssec:mag_indep_selection}

\begin{figure}
  \begin{center}
    \plotone{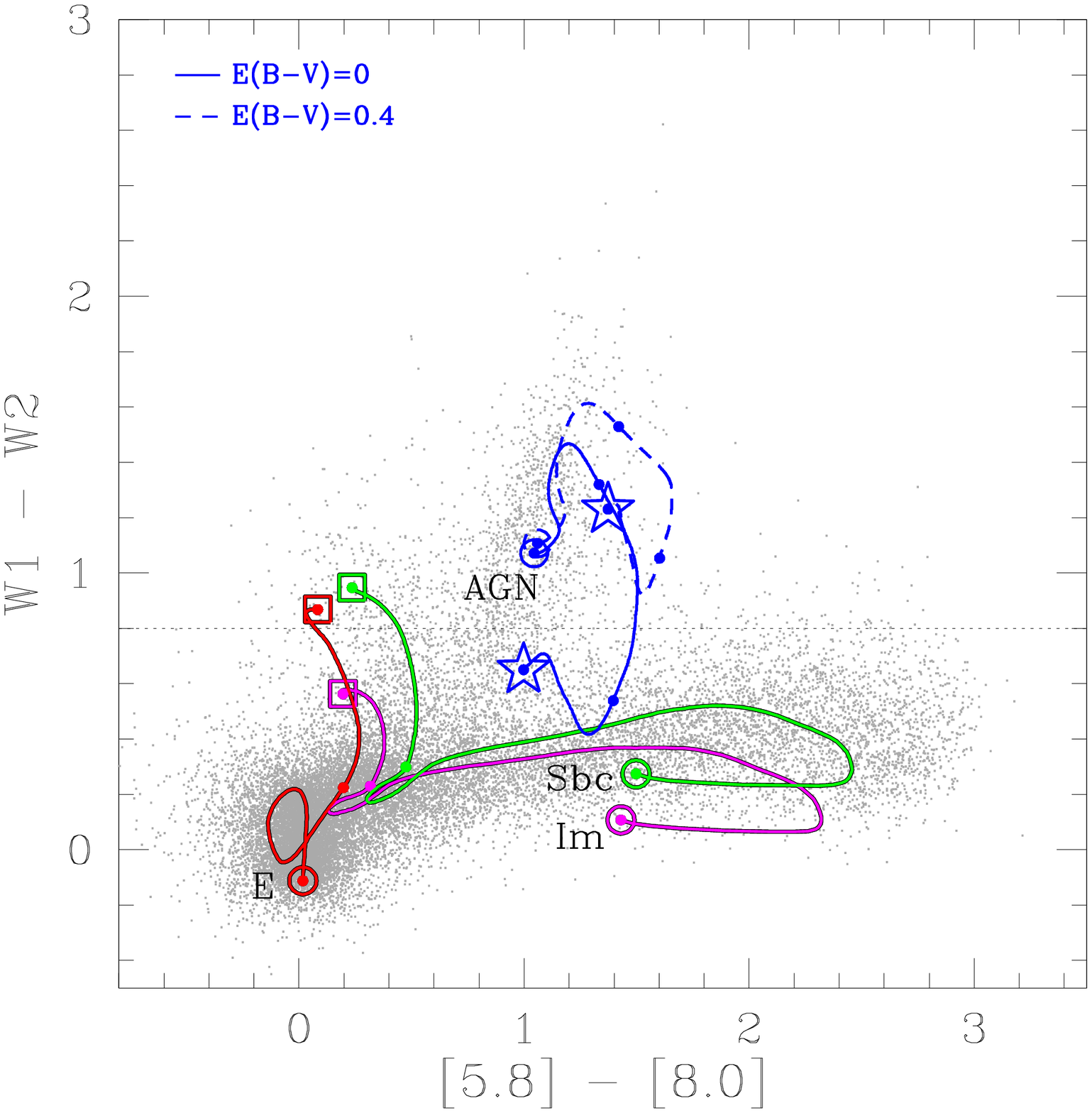}
    \caption{Distribution of [5.8]--[8.0] and W1--W2 colors in the
      Bo\"otes field. The lines show the colors of the galaxy and AGN
      SED templates of \citet{assef10}. The color of the galaxy
      templates E ({\it{red line}}), Sbc ({\it{green line}}) and Im
      ({\it{magenta line}}) are shown between redshifts 0 ({\it{open
          circle}}) and 2 ({\it{open square}}), with dots in the
      tracks in steps of $\Delta z = 1$. The AGN template is shown
      without reddening ({\it{solid blue line}}) and with $E(B-V)=0.4$
      ({\it{dashed blue line}}), in the redshift interval between
      $z=0$ ({\it{open circle}}) and $z=6$ ({\it{open star}}). Dots
      in the AGN color tracks are spaced by $\Delta z = 2$. The gray
      dots show all the WISE sources in the NDWFS field with
      $W2<15.73$.}
    \label{fg:wise_cmd_tracks}
  \end{center}
\end{figure}

In \citetalias{stern12} we investigated the distribution of quasars in
WISE color space in the COSMOS survey field. We found that for objects
with W2$<$15.05~mag (W2 $S/N \geq 10$ at that ecliptic latitude), the
simple color cut based on the two shortest wavelength WISE bands
\begin{equation}\label{eq:stern12_sel}
  \rm W1 - \rm W2\ \geq\ 0.8,
\end{equation}
\noindent provides an effective criterion to separate AGN from
inactive galaxies. When compared to the IRAC color selection method of
\citet[][see \citealt{assef10} for discussion about its
  reliability]{stern05}, the WISE color criterion selection recovers
78\% of the IRAC-selected AGN with a 95\% reliability. Notably, six of
the AGN candidates selected by IRAC and WISE were not detected in the
1.8~Ms {\it{Chandra}} survey of the COSMOS field
\citep[C-COSMOS;][]{elvis09}, suggesting they may be Compton-thick
(see \citetalias{stern12} for details). The reason behind the success
of this criterion is clearly illustrated in Figure
\ref{fg:wise_cmd_tracks}. This figure shows mid-IR color as a function
of redshift for the AGN and galaxy SED templates of
\citet{assef10}. For $z\lesssim 3$, the W1--W2 color of unobscured AGN
is well above the color cut of 0.8~mag. At higher redshift, reddened
AGN ($E(B-V)\gtrsim 0.4$) can also be redder than this color cut. In
practice, however, it is exceedingly uncommon to find high redshift,
highly reddened AGN bright enough to be detected by WISE and so much
more luminous that their host galaxy to dominate the rest-frame
optical emission. Indeed, in \citetalias{stern12} we found no quasars
with $z>3$ in the sample. We also find very few galaxies with
$z\gtrsim 1$, since the WISE observations of the COSMOS field are not
deep enough to find normal galaxies at high redshifts. While other
populations such as some ULIRGs and brown dwarfs can have even redder
W1--W2 colors, they are too rare in comparison to AGN to be a
significant source of contamination. Note that Figure
\ref{fg:wise_cmd_tracks} implies that this W1--W2 color selection is
biased against AGN which are faint with respect to their host
galaxies. If the flux in the WISE bands is dominated by the galaxy,
the colors will drop below the selection limit, moving towards the
galaxy locus at $\rm W1 - \rm W2 \sim 0$. We discuss this further in
\S\ref{ssec:sed_analisys}.

The criteria of \citetalias{stern12} are readily applicable to the
all-sky WISE survey, and are demonstrated to be both reliable and
complete to the shallow depth of the WISE observations of the COSMOS
field. However, due to the limited size of that field, it does not
have the statistical power to address many interesting and pressing
issues in AGN studies, such as AGN evolution, accretion rates, and
dust distributions. More importantly, since most of the WISE survey
area has deeper coverage than in the COSMOS field, alternative
selection criteria are valuable for a census of WISE-selected AGN in
these deeper regions. \citet{jarrett11} has shown that in the deepest
WISE fields at the ecliptic poles, where W1 and W2 are confusion
limited, the addition of W3 is a very useful aid in the identification
of AGN. Our intention is to bridge these two extremes, proposing a
robust WISE AGN selection technique for fields with intermediate
depth.

In order to extend the study of \citetalias{stern12}, we turn to the
NDWFS Bo\"otes field, which helps with both issues highlighted above:
it has a WISE median coverage of 30 frames, almost three times that of
COSMOS, and extends over $9~\rm deg^2$, an area $4.5$ times larger. We
start by replicating the selection criterion of \citetalias{stern12}
in the Bo\"otes field, but up to the W2 10$\sigma$ depth provided by
the full co-added data. While this implies a sample $0.68~\rm mag$
deeper, it maintains the error properties of the sample and so
provides a meaningful comparison.

\begin{figure}
  \begin{center}
    \plotone{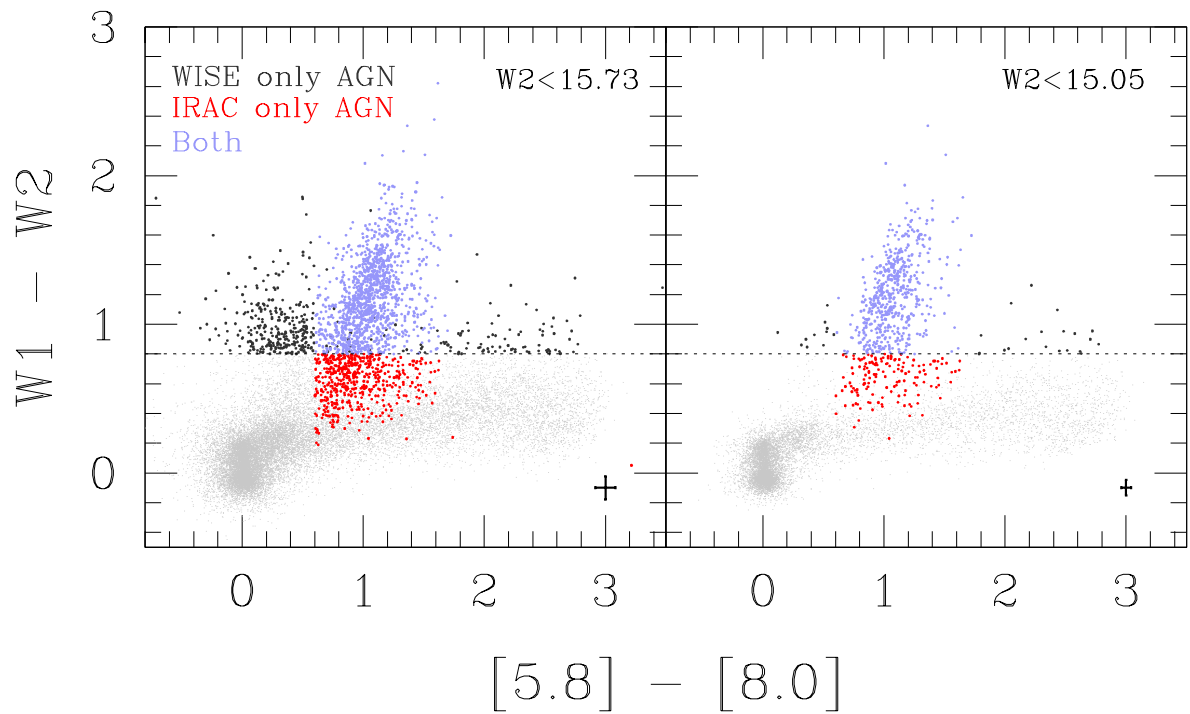}
    \caption{WISE W1--W2 versus SDWFS [5.8]--[8.0] colors for WISE
      sources in the NDWFS Bo\"otes field. The left panel shows
      sources with $\rm W2<15.73$, the $10\sigma$ WISE detection limit
      in the Bo\"otes field, while the right panel shows sources
      limited to $\rm W2<15.05$, corresponding to the W2 $S/N>10$
      limit in the COSMOS field. Objects are separated into non-AGN
      candidates ({\it{light gray dots}}), WISE and IRAC AGN
      candidates ({\it{blue dots}}), WISE-only candidates ({\it{black
          dots}}) and IRAC-only candidates ({\it{red dots}}). The
      median photometric uncertainty for each sample is shown in the
      lower right corner of each panel.}
    \label{fg:wise_cmd}
  \end{center}
\end{figure}

The left panel of Figure \ref{fg:wise_cmd} shows the WISE color
distribution of sources in our sample. Comparing to sources selected
as AGN by their IRAC colors \citep{stern05}, we find that a simple
W1$-$W2$\geq$0.8 color cut identifies 70\% of the IRAC-selected AGN
with 70\% reliability. Compared to applying this cut at the 10$\sigma$
WISE depth of COSMOS as reported in \citetalias{stern12}, the drop in
completeness is relatively small, from 78\% to 70\%. The decrease in
reliability from 95\% to 70\% is very significant, however, and is
simply due to the modest increase in field depth. If we limit the
Bo\"otes field analysis to the 10$\sigma$ W2 level of the COSMOS field
(W2$<$15.05, right panel of Fig. \ref{fg:wise_cmd}), we recover
similar AGN demographics to that reported in \citetalias{stern12},
with 78\% completeness and 94\% reliability.

The lower completeness means an increase in IRAC-selected AGN detected
but not identified by the simple WISE color criterion. This is likely
due to a combination of (i) a small number of $z\gtrsim 3$ Type 1 AGN,
which are known to be excluded by the \citetalias{stern12} color
selection; (ii) low-redshift, low-luminosity AGN with hosts bright
enough to move their mid-IR colors below the WISE selection limit, but
red enough to be picked by the \citet{stern05} IRAC selection
criteria; and (iii) a higher incidence of contamination by $z\sim 0.5$
star-forming galaxies to the IRAC selection criterion which
artificially lowers the completeness --- though we note that this
contamination is expected to be small at the depth of SDWFS
\citep[see][]{assef10,donley12}.

The cause of the significantly lower reliability obtained in the
Bo\"otes field compared to the COSMOS field is readily apparent in
Figure \ref{fg:wise_cmd}. The modestly deeper WISE sample increases
the number of contaminating galaxies, particularly to the left of the
QSO locus. These correspond to high-redshift ($z\sim 1-1.5$)
galaxies. The observed W2 magnitude of $z\sim 1-1.5$ $L_{*}$ galaxies
evolves very slowly with redshift \citep[see, e.g., Fig. 1
  of][]{eisenhardt08}, so there is a huge increase in contamination as
soon as the W1 magnitude limit is deep enough to begin including these
galaxies. In the \citet{stern05} IRAC selection criterion this problem
is controlled using the [5.8]--[8.0] color, but the longer wavelength
WISE bands are too shallow to help.

\subsection{Magnitude-Dependent AGN Color-Selection}\label{ssec:mag_dep_cut}

It is apparent from Figure \ref{fg:wise_cmd} that an improved method
to select AGN may be possible if we allow our color cut to evolve with
magnitude since the major contaminants are either low-redshift, nearby
star-forming galaxies which are intrinsically faint, or high-redshift,
passive galaxies that are luminous enough to be bright in the WISE
bands. In order to design a magnitude-dependent AGN color selection
method that is applicable over the whole sky, we will go to fainter
WISE fluxes than afforded by the 10$\sigma$ W2 limit in the Bo\"otes
field. Note, however, that as we go to fainter W2 magnitudes, it
becomes unreliable to use an AGN control sample based on the IRAC
color criteria of \citet{stern05}, as it is susceptible to
contamination by high-redshift galaxies once the errors in the SDWFS
IRAC [5.8] and [8.0] fluxes become too large \citep[see,
  e.g.,][]{donley12}. Instead, we define the control sample as all
objects whose best-fit UV--mid-IR SEDs have a strong AGN component, as
indicated by requiring $\hat{a}>0.5$. For significantly lower levels
of AGN activity, it becomes necessary to differentiate between objects
where the AGN component of the fit is real and when it has only been
used to mathematically improve the $\chi^2$ to accommodate lower
quality photometry or mimic a galaxy component missing from the
templates. This falls beyond the scope of the current work, and a full
analysis on this topic is presented by
\citet{chung12}. \citet{assef10} has shown that the \citet{stern05}
criterion is biased towards objects with large $\hat{a}$ values (see
also \S\ref{ssec:sed_analisys}), so we are not considerably changing
the physical properties of the control sample by using this
definition.

\begin{figure}
  \begin{center}
    \plotone{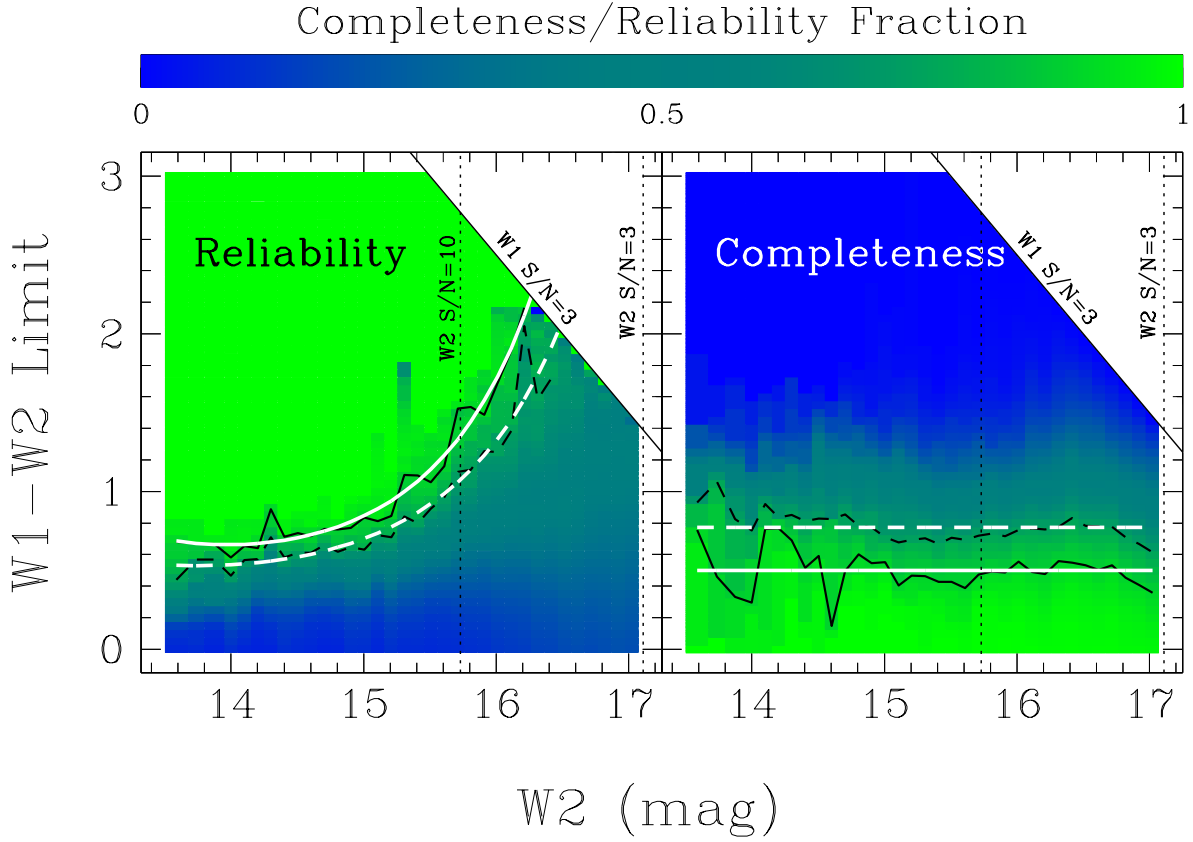}
    \caption{Reliability ({\it{left panel}}) and completeness
      ({\it{right panel}}) of AGN candidates defined by $\hat{a}>0.5$
      selected by a color cut on $\rm W1-\rm W2$ as a function of $\rm
      W2$ magnitude. Reliability and completeness of 90\% (75\%) are
      shown as a function of magnitude by the solid (dashed) black
      lines. Objects redder than the top right corner of the panels
      are missing due to the W1 $S/N > 3$ requirement. The proposed
      reliability-optimized criteria (eqn.[\ref{eq:rel_sel}]) for 90\%
      ($R_{90}$) and 75\% ($R_{75}$) reliability are shown in the left
      panel by the white solid and dashed lines, respectively. The
      completeness-optimized criteria (eqn.[\ref{eq:comp_sel}]) for
      90\% ($C_{90}$) and 75\% ($C_{75}$) completeness are shown in
      the right panel with the same respective line styles.}
    \label{fg:mag_color_cut_sed}
  \end{center}
\end{figure}

Figure \ref{fg:mag_color_cut_sed} shows the completeness and
reliability obtained as a function of W2 magnitude and the minimum
W1--W2 color limit adopted to select AGN. We have required a minimum
detection threshold of 3$\sigma$ for W1 in order to have a reasonably
precise WISE color. At bright W2, a color cut of 0.6 is sufficient to
obtain high reliability and high completeness. Towards fainter W2
magnitudes, high reliability requires redder color cuts in order to
remove contaminating galaxies, which also leads to lower
completeness. The completeness of a color cut is relatively
independent of W2 magnitude.

\begin{figure}
  \begin{center}
    \plotone{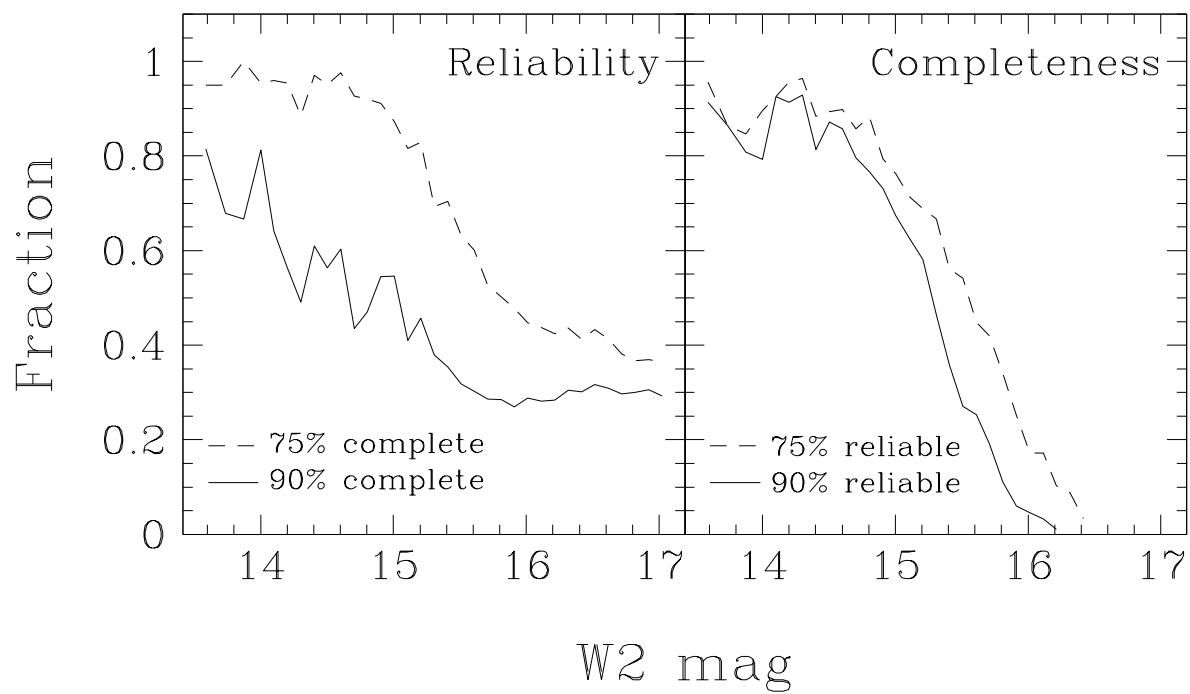}
    \caption{The left panel shows the reliability as a function of W2
      magnitude for the 90\% (solid line) and 75\% (dashed line)
      completeness-optimized AGN selection criteria. The right panel
      shows the completeness as a function of W2 magnitude for the
      90\% (solid line) and 75\% (dashed line) reliability-optimized
      AGN selection criteria.}
      \label{fg:rel_comp_fixed_opposite}
  \end{center}
\end{figure}

Figure \ref{fg:mag_color_cut_sed} shows the bluest W1--W2 color at
which 90\% and 75\% reliability is reached for a given W2
magnitude. While there is significant noise in these curves, they are
reasonably well described by an exponential in $\rm W2^2$. Hence, we
propose a WISE AGN color selection limit optimized for reliability
given by
\begin{equation}\label{eq:rel_sel}
\rm W1 - \rm W2 > \alpha_R\ \exp\left\{\beta_R\ (\rm
W2-\gamma_{\rm R})^2\right\}.
\end{equation}
\noindent For W2$<$17.11, we achieve a reliability of $\sim$90\% with
$(\alpha_{R90}, \beta_{R90}, \gamma_{R90}) = (0.662, 0.232, 13.97)$.
The corresponding values for a reliability of $\sim$75\% are
$(\alpha_{R75}, \beta_{R75}, \gamma_{R75}) = (0.530, 0.183, 13.76)$.
The 90\% (75\%) reliability criterion reach our imposed W1 $S/N > 3$
limit at a W2 magnitude of 16.26 (16.45). The right panel of Figure
\ref{fg:rel_comp_fixed_opposite} shows the completeness as a function
of W2 magnitude for each of the criteria. Only considering objects
brighter than this limit in W2, the 90\% reliability criterion
identifies 1174 AGN candidates, of which 1060 (90\%) have their
bolometric luminosities dominated by the AGN (e.g., $ \hat{a}> 0.5$).
For the 75\% reliability criterion, we identify 2306 AGN candidates,
of which 1752 (76\%) are AGN-dominated.  At a shallower depth of W2
$<$15.73, corresponding to $S/N > 10$ in the Bo\"otes field, the 90\%
reliability curve identifies 1051 AGN candidates of which 950 (90\%)
are AGN-dominated, while the 75\% reliability curve identifies 1582
AGN candidates of which 1200 (76\%) are AGN-dominated. For comparison,
a simple W1--W2$\geq$0.8 cut (e.g., similar to the criterion of
\citetalias{stern12}, but without its magnitude cut) finds 1746 AGN
candidates with 74\% reliability at a depth of W2 $= 15.73$.  This
census increases to 17997 AGN candidates to a depth of W2 $= 17.11$,
albeit with a reliability that drops to 45\%. These statistics are
summarized in Tables \ref{tab:dens_sn3} and \ref{tab:dens_sn10}. In
the next section (\S\ref{ssec:lit_comp}) we compare this new
magnitude-dependent AGN selection criterion to several other WISE AGN
selection criteria that have recently been proposed in the literature.

Figure \ref{fg:mag_color_cut_sed} also shows the reddest W1--W2 color
at which 90\% and 75\% completeness is reached for objects with
bolometric luminosities dominated by the AGN emission. In this case, a
reasonable description of the completeness boundary is given by the
magnitude independent color cut
\begin{equation}\label{eq:comp_sel}
\rm W1 - \rm W2 > \delta_C,
\end{equation}
\noindent where $\delta_{C90} = 0.50$ for 90\% completeness and
$\delta_{C75} = 0.77$ for 75\% completeness. Note that the 75\%
completeness criterion is basically equivalent to the cut proposed by
\citetalias{stern12}, shown in equation \ref{eq:stern12_sel}, but
without the flux cut. The left panel of Figure
\ref{fg:rel_comp_fixed_opposite} shows the reliability as a function
of W2 magnitude for each of the criteria. It is important to stress
that these criteria are appropriate only for strong AGN with respect
to their hosts due to the $\hat{a}>0.5$ requirement.

While the magnitude dependence in the reliability-optimized criterion
is caused in part by the much higher number of high redshift galaxies
at fainter fluxes (see discussion in
\S\ref{ssec:mag_indep_selection}), it is also driven by the
increasingly large errors in W1 and W2 at lower $S/N$. Since the $S/N$
of WISE observations varies significantly across the sky, in principle
$\alpha_R$, $\beta_R$ and $\gamma_R$ may depend on ecliptic
latitude. To test the strength of this dependence, we simulate the
distribution of W1--W2 and W2 magnitudes for different WISE field
depths and estimate $\alpha_R$, $\beta_R$ and $\gamma_R$ in each
case. For this we use the magnitudes obtained from the SED modeling of
every object and we approximate that the $S/N$ for a flux $F$ depends
on field depth as
\begin{equation}
  \frac{S}{N}\ =\ K_1\ \sqrt{\frac{N_F}{30}}\ \frac{F}{\sqrt{F +
      F_{\rm Sky}}},
\end{equation}
\noindent where $N_F$ is the number of individual WISE 11s frames used
to build the catalog image, $K_1$ is a constant and $F_{\rm Sky}$ is
the background flux. This formulation neglects the effects of
confusion as well as the variation of $F_{\rm Sky}$ with sky position,
which is not uniform across the sky. However, our approximation should
give a good general idea of how the parameters in question vary with
$N_F$. As discussed earlier, $N_F^{\rm Bootes} = 30$, and we use the
Bo\"otes data to fit for $K_1$ and $F_{\rm Sky}$. We find that from
$N_F=10$ to 50, no significant variation is observed for the
$\alpha_{\rm R}$ and $\gamma_{\rm R}$ parameters for both the 90\% and
75\% criteria. The parameter $\beta_R$ is observed to decrease
linearly by a factor of $\sim4$ between $N_F=8$ and $N_F=25$, and is
approximately independent of the depth of the field for $N_F\gtrsim
25$. Considering then a modified $\beta_R^{\prime} = \beta_R^{{\rm
    Boo}}\ (5.41-0.176 N_F)$ may be necessary to achieve the proposed
reliability levels in WISE fields with $N_F<25$, where
$\beta_R^{{\rm{Boo}}}$ is the value obtained for the Bo\"otes
field. We also repeat the experiment for the completeness optimized
criteria, and find that in the same range of $N_F$, $\delta_C$ is
approximately constant.

We refer to the reliability-optimized selection as $R_{90}$ and
$R_{75}$ for 90\% and 75\% reliability, respectively, while $C_{90}$
and $C_{75}$ refer to the completeness-optimized criteria.  Whether
scientific interest lies in maximizing completeness or reliability
depends on the problem at hand. However, it will be most common to
wish to maximize reliability, so in the next sections we will focus on
results for the highest reliability selection.

\subsection{Comparison with the Literature}\label{ssec:lit_comp}

AGN identification using mid-IR broad-band photometry is now a well
studied problem. The first classification schemes were developed for
{\it{Spitzer}} IRAC and MIPS photometry \citep[see,
  e.g.,][]{lacy04,lacy07,stern05,alonsoherrero06,messias12} and have
been shown to be very successful in terms of both reliability and
completeness. In the previous section we developed four AGN
identification schemes using WISE W1 and W2 photometry, optimized to
produce samples with different levels of either reliability or
completeness. Several other WISE criteria have also recently been
developed. Here we briefly discuss several of these criteria and
discuss how they compare to our selection criteria. This is meant to
be an illustrative rather than an exhaustive exercise; we do not
discuss all the published mid-IR selection techniques.

Tables \ref{tab:dens_sn3}, \ref{tab:dens_sn10} and
\ref{tab:dens_cosmos} show the surface density of AGN candidates,
their reliability and their completeness for the selection criteria of
\S\ref{ssec:mag_dep_cut}, \citetalias{stern12}, \citet{jarrett11},
\citet{mateos12} and \citet{wu12}. Table \ref{tab:dens_sn3} restricts
the samples to W2$<$17.11 (W2 $S/N>3$ in Bo\"otes), while Table
\ref{tab:dens_sn10} uses the more restrictive flux cut of W2$<$15.73
(W2 $S/N>10$ in Bo\"otes). The samples used in all tables are also
restricted to W1$<$18.50, but no restriction is applied in W3 and
W4. For completeness, Table \ref{tab:dens_cosmos} further limits the
sample to W2$<$15.05 (W2 $S/N>10$ in COSMOS), which is representative
of the shallowest WISE observations, but we will not discuss it in
detail. We also include the selection criteria proposed by
\citet{assef10}, which was obtained by simulating WISE photometry
using SED models of all objects in SDWFS. For comparison, we also show
the numbers for the IRAC-based selection criteria of
\citet{lacy04,lacy07}, \citet{stern05} and \citet{messias12},
calculated using the SDWFS photometry. As was done in the previous
section, reliability and completeness are measured against the number
of objects whose SED fits have $\hat{a}>0.5$.

For the WISE selection methods, regardless of the W2 depth, the most
reliable sample is that based on the W1, W2, W3 and W4 selection
criteria of \citet{assef10}, with a 97\% and 98\% reliability for
W2$<$17.11 and W2$<$15.73, respectively. However, because it requires
that W4 is detected, it also has the lowest completeness (3\% and 19\%
for W2$<$17.11 and W2$<$15.73, respectively) as measured by the
surface density of AGN candidates, with only 44~deg$^{-2}$ candidates
with W2$<$17.11. Our $R_{90}$ is the second most reliable criterion,
with 90\% reliability by design, but it has a much higher
completeness, with 53\% for W2$<$15.73 and 9\% for W2$<$17.11, which
translates into AGN candidate surfaces densities of 117 and
130~deg$^{-2}$ respectively. The four-band criterion of
\citet{mateos12} also has high reliability, although it is below our
$R_{90}$ criterion in both reliability and completeness. Both of the
W1, W2 and W3 based selection criteria of \citet{jarrett11} and
\citet{mateos12} are similar in reliability and completeness,
comparable to our $R_{75}$ criterion for W2$<$15.73, but somewhat less
reliable for W2$<$17.11. As discussed by \citet{jarrett11}, the
strength of these criteria are in the deepest WISE fields, where the
W1 and W2 depths are below the confusion limit.

With respect to the IRAC-based criteria, the most reliable of those
shown are the criteria of \citet{messias12}, followed by that of
\citet{stern05}. In terms of completeness levels, we note that all
criteria shown are similar, except for the highest reliability ``KIM''
criteria of \citet{messias12} based on $K_s$, [4.5], [8.0] and MIPS
24$\mu$m photometry, which has a lower completeness of 22\% for
W2$<$17.11 and 51\% for W2$<$15.73. It is also important to notice
that, in principle, completeness and reliability could be improved by
further adding more information based on other wavelength
regimes. Such is the case, for example, with the ``$S_{IX}$''
selection scheme of \citet{edelson12}, which combines WISE, 2MASS and
ROSAT data to identify the brightest AGN in sky. Including this kind
of selection is, however, beyond the scope of this comparison.

\section{Properties of WISE AGN Candidates}\label{sec:wise_agn_sed_properties}

In this section we study the properties of the WISE AGN candidates
selected using the criteria developed in the previous section. We
first discuss the accuracy to which we can determine photometric
redshifts for them. In \S\ref{ssec:redshift_dist} we discuss their
redshift distribution, and in \S\ref{ssec:sed_analisys} we discuss the
parameters derived from our SED fitting. In \S\ref{ssec:keck} we
present spectroscopic observations of a sample of photometrically
selected high redshift Type 2 AGN candidates.

\subsection{Photometric Redshift Accuracy for WISE AGN Candidates}\label{ssec:photoz_acc}

Several authors
\citep[e.g.,][]{brodwin06,rowan-robinson08,salvato09,assef10} have
shown that photometric redshifts of Type 1 AGN are relatively
inaccurate when relying solely on broad-band photometry, as is our
case. This is mostly due to the lack of strong spectral features that
are necessary for anchoring the photometric redshift estimates. Our
AGN sample is, however, brighter than those typically studied for
photometric redshifts, and has a considerable number of Type 2
AGN. Photometric redshifts for Type 2 AGN may be better because the
spectral features of the host galaxy are relatively stronger.

We estimate photometric redshifts as discussed in
\S\ref{ssec:sed_fits}, using, in addition to WISE, all the UV through
mid-IR broad-band photometry of the field described in
\S\ref{ssec:ndwfs}. As in \citet{assef10}, we quantify the photometric
redshift accuracy using the statistic
\begin{equation}
  \Delta z\ =\ \left[ \frac{1}{N}\ \sum_i \left(\frac{z_{\rm phot}^i -
      z_{\rm spec}^i}{1+z_{\rm spec}^i}\right)^2 \right]^{1/2},
\end{equation}
\noindent where the index $i$ sums over all objects in a sample and
$N$ is their total number. This estimate of the dispersion, however,
is typically driven by outliers, so we also estimate $\Delta z_{95}$,
the dispersion calculated including only the 95\% of objects with the
photometric redshift estimates closest to the spectroscopic estimate.

\begin{figure}
  \begin{center}
    \plotone{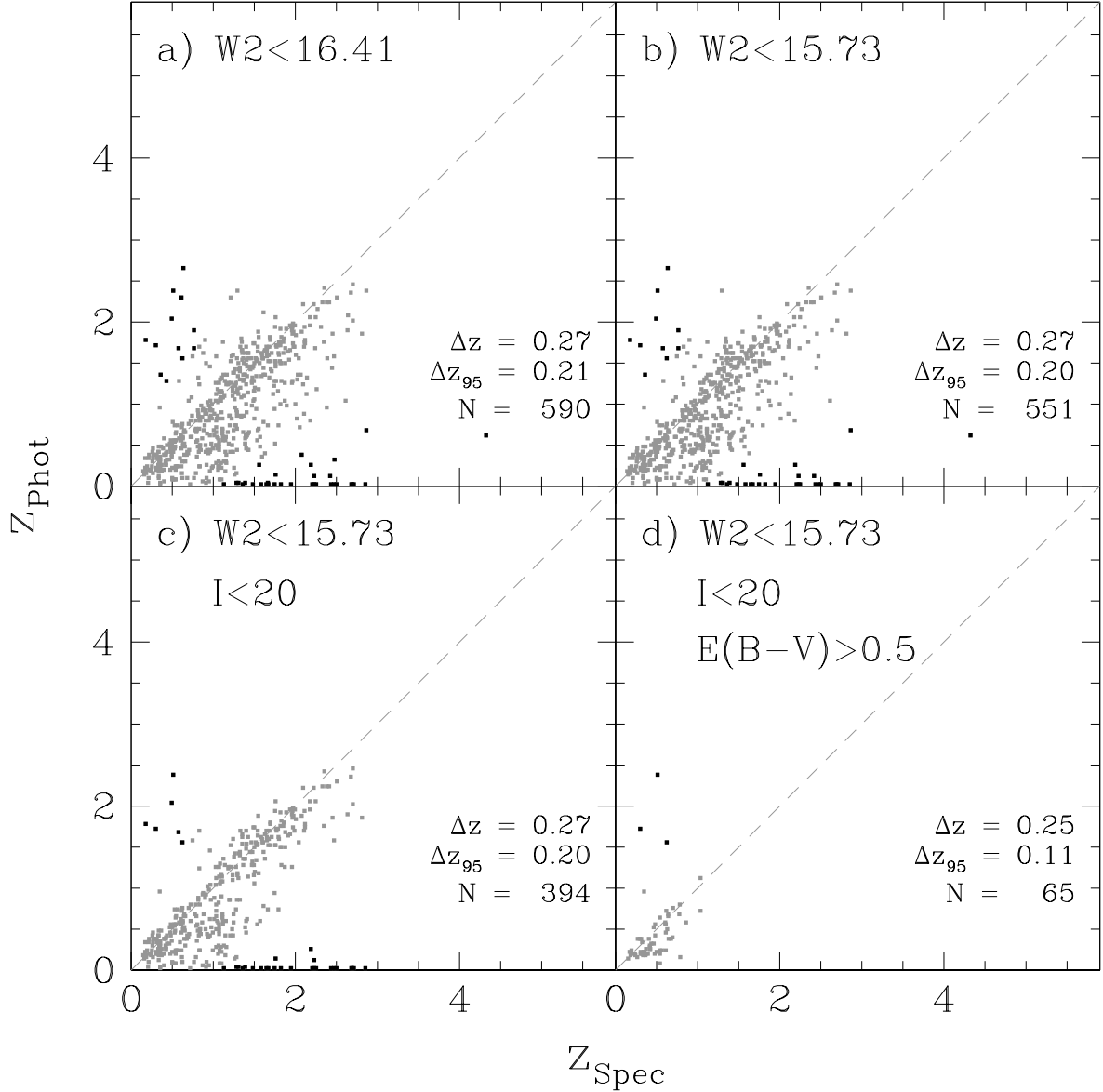}
    \caption{Comparison of photometric and spectroscopic redshifts for
      the $R_{90}$ sample of AGN candidates for: {\it{a)}} the full W2
      depth sample; {\it{b)}} limited to objects with $\rm W2<15.73$;
      {\it{c)}} further limited to objects with $I<20$; and {\it{d)}}
      even further limited to objects $E(B-V)>0.5$. Each panel shows
      the dispersion between the photometric and spectroscopic
      redshifts for the full sample ($\Delta z$) and for the 95\%
      objects with the best estimates to minimize the effect of
      outliers ($\Delta z_{95}$). Black points correspond to objects
      with $|(z_{\rm phot}-z_{\rm spec})/(1+z_{\rm spec})| > 0.5$.}
    \label{fg:dz_r90}
  \end{center}
\end{figure}

Panel a) of Figure \ref{fg:dz_r90} shows the spectroscopic and
photometric redshifts obtained for the full W2 depth $R_{90}$ AGN
candidates, limited to objects with $\hat{a}>0.5$ to be certain we
only study the objects of interest. Table \ref{tab:photoz} shows the
dispersion as well as the median offsets for the remaining criteria,
again limited to $\hat{a}>0.5$. It also shows the number of AGN used
to compute the statistic and the fraction of objects in every
selection criteria that have spectroscopic redshifts. Irrespective of
the selection method, the photometric redshifts are fairly inaccurate,
with $\Delta z_{95} = 0.20-0.23$ ($\Delta z = 0.27-0.31$). This is
consistent with the results presented by \citet{assef10} for a
similar, but fainter, sample of objects. The pile-up of objects at
very low $z_{\rm phot}$ is a degeneracy caused by the galaxy
luminosity prior. However, these are only a small part of the sample,
and eliminating the prior results in even less accurate estimates for
the general population. Panel b) shows that little is gained in terms
of the accuracy when limiting the sample to the brighter W2 $S/N>10$
objects. The same is observed when the sample is further limited by
requiring $I<20$, as shown in Panel c).

Considering that photometric redshift estimates for Type 2 AGN may be
more accurate (see above), we further split the bright, final sample
(${\rm W2}<15.73~{\rm and}~I<20$ and $\hat{a}>0.5$) and only
investigate objects with considerable obscuration, $E(B-V) > 0.5$. We
find that $\Delta z_{95}$ drops by $\sim45\%$, although $\Delta z$
either decreases only slightly ($R_{90}$ and $C_{75}$), or increases
($R_{75}$ and $C_{90}$). That these accuracies are still much worse
than the $\Delta z_{95}\sim 0.04$ found by \citet{assef10} for
galaxies of equivalent brightness is most likely due to the sample
requirement that $\hat{a}>0.5$, meaning that even though reddened, the
accretion disk emission is still likely dominant, or at least
significant, in many of the broad-bands used.

As mentioned earlier, \citet{assef10} showed that although photometric
redshifts for AGN based on broad-bands can be inaccurate, the value of
$\hat{a}$ obtained from the corresponding SED fit is insensitive to
the redshift accuracy, i.e., $\hat{a}$ is relatively independent of
photometric redshift. As one of our goals is to study the obscuration
in AGN in a statistically significant manner, we can ask if this holds
for the inferred reddening of the AGN component. Hence, we compare the
estimates of $E(B-V)$ obtained from the SED fits using the photometric
redshift and the spectroscopic redshift estimates. We find that for
objects where there is good agreement between $z_{\rm phot}$ and
$z_{\rm spec}$, the two estimates of $E(B-V)$ are consistent with each
other. Unfortunately, however, when the redshift estimates disagree,
so do the AGN obscuration estimates, with systematically low $E(B-V)$
values when assuming $z=z_{\rm phot}$. This will be of particular
importance in \S\ref{sec:dust_properties}.

\subsection{Redshift Distribution of WISE AGN Candidates}\label{ssec:redshift_dist}

Using the cuts developed in the previous section we now study the
redshift distribution of the different samples of AGN
candidates. Although we have a large amount of spectroscopic
observations in the Bo\"otes field, we are still missing spectroscopic
redshifts for a considerable number of our AGN candidates (see Table
\ref{tab:photoz} for details). For the objects without spectroscopic
redshifts, we use the photometric redshift estimates detailed in
\S\ref{ssec:sed_fits}, although these may not be very accurate (see
\S\ref{ssec:photoz_acc}). We focus on the $R_{90}$ sample, which
mitigates this issue as these objects are the ones most likely to have
spectra from the AGES survey \citep[see \S\ref{ssec:ndwfs} and][for
  details]{kochanek12}.

\begin{figure}
  \begin{center}
    \plotone{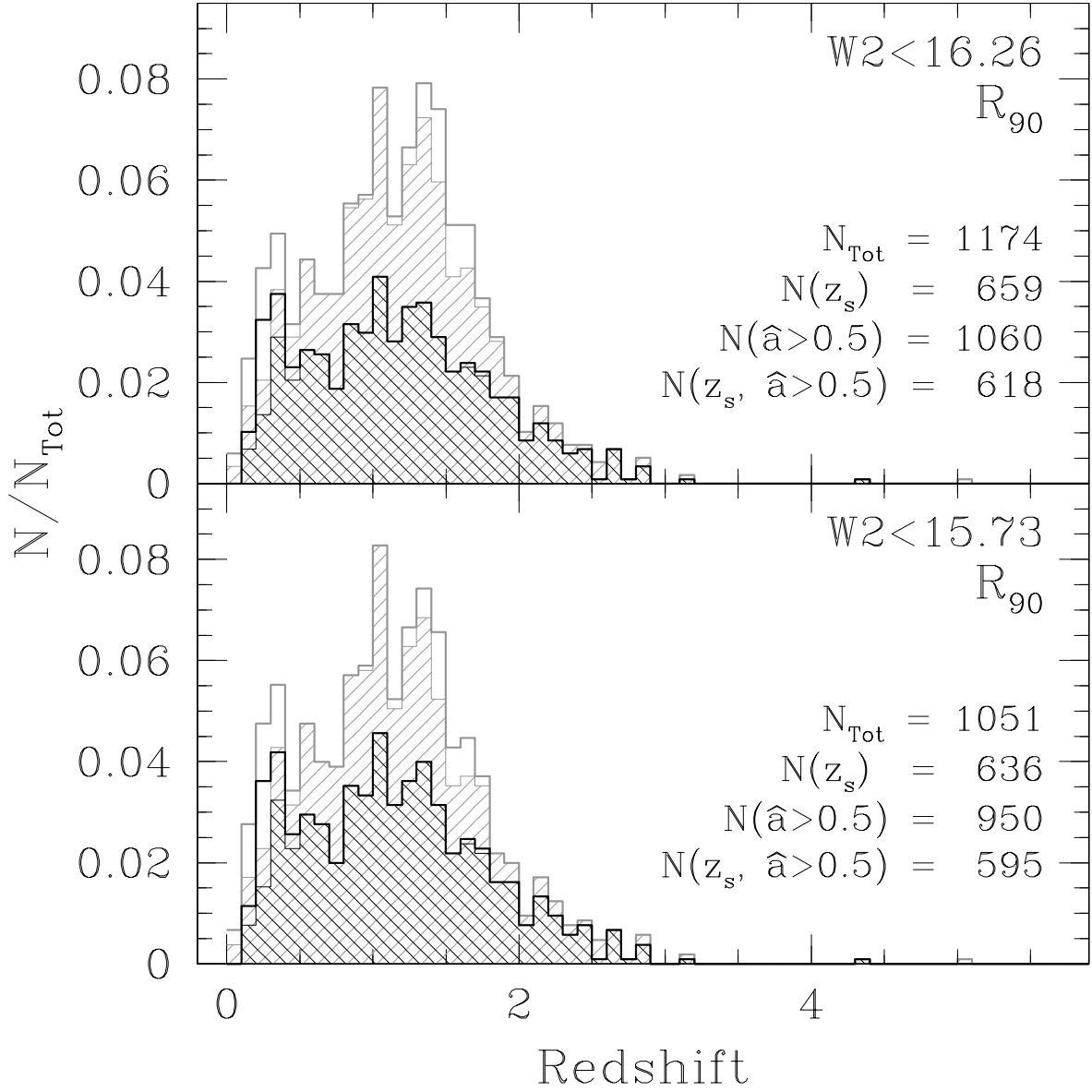}
    \caption{Redshift distribution of the $R_{90}$ sample of AGN
      candidates for the full W2 depth of the field ({\it{top panel}})
      and limited to objects with $\rm W2<15.73$ ({\it{bottom
          panel}}). Black histograms show objects with spectroscopic
      redshifts, while gray histograms add objects with photometric
      redshift estimates. Shaded histograms only include objects with
      $\hat{a}>0.5$ while open histograms use objects with all
      $\hat{a}$ values, including the contaminants.}
    \label{fg:redshift_r90}
  \end{center}
\end{figure}

Figure \ref{fg:redshift_r90} shows the redshift distribution of the
W2, W1 $S/N>3$ depth $R_{90}$ sample. The resulting distribution of
AGN is double peaked, with the main peak at $1 \lesssim z \lesssim 2$
and a smaller peak at $z\sim 0.25$. Almost no objects are at $z\gtrsim
3$. This distribution reflects that WISE has a high sensitivity to
obscured AGN at lower redshifts, where the AGN emission still
dominates the observed W1 and W2 fluxes. However, as the redshift
increases or the galaxy host contributions become larger, the bias
against obscured sources increases. This causes the minimum at $z\sim
0.75$, followed by an increase simply from the increase in comoving
volume probed. The W1--W2 color of Type 1 AGN is reddest at $1
\lesssim z \lesssim 2$ (see Fig. \ref{fg:wise_cmd_tracks}), and
progressively gets bluer at higher redshift, falling completely out of
the selection criteria by $z\sim 3$. A similar behavior is observed
for the $R_{75}$, $C_{75}$ and $C_{90}$ samples, although since
contaminants appear preferentially at high redshifts (see
\S\ref{ssec:mag_dep_cut}), the balance between the peaks for the
complete $\hat{a}$ samples is modified. Figure \ref{fg:redshift_r90}
shows that a large number of the $R_{90}$ AGN candidates (56\%) have
spectroscopic redshifts. Furthermore, objects lacking spectroscopic
redshifts tend to follow a similar photometric redshift distribution,
implying that although the uncertainties in the photometric redshifts
are very large, they do not seem to systematically bias the
distribution. Limiting the samples to only the brighter W2 $S/N>10$
objects does not significantly change the shape of the redshift
distribution.

\subsection{SED Analysis of WISE AGN Candidates}\label{ssec:sed_analisys}

\begin{figure}
  \begin{center}
    \plotone{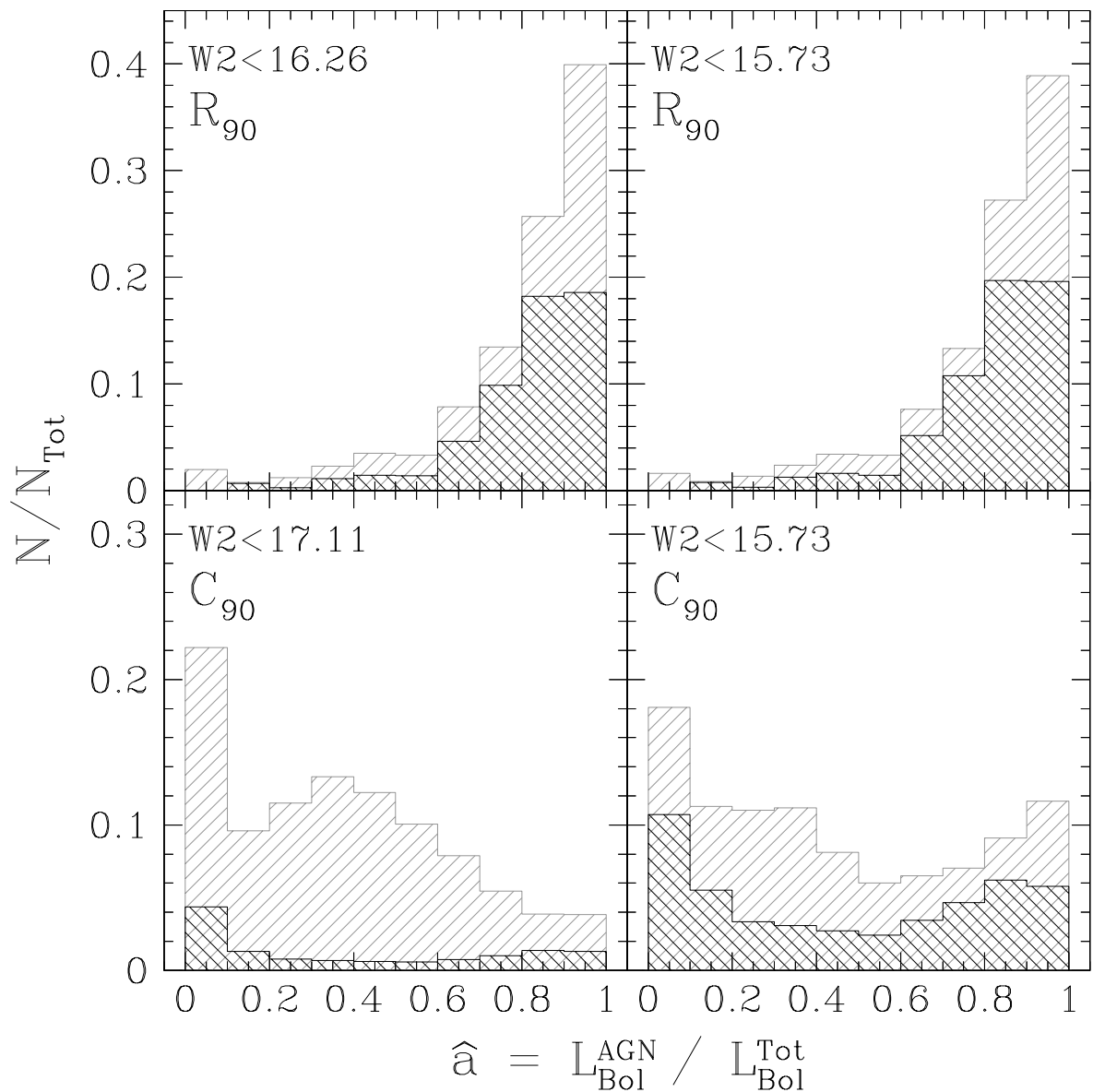}
    \caption{Distribution of $\hat{a}$ values for our $R_{90}$
      ({\it{top}}) and $C_{90}$ ({\it{bottom}}) AGN candidate
      samples. Left panels show the full W2 depth samples, while the
      right panels are limited to objects with $\rm W2<15.73$. Black
      histograms include only objects with spectroscopic redshifts,
      while gray histograms also include objects with photometric
      redshifts.}
    \label{fg:ahat_r90_c90}
  \end{center}
\end{figure}

A simple way of quantifying the contamination rates in the criteria we
have defined is by looking at the best-fit combination of SED
templates to their photometry. The most relevant parameter is
$\hat{a}$, defined in equation (\ref{eq:ahat}). As mentioned earlier,
this parameter has the useful property of being relatively insensitive
to photometric redshift uncertainties \citep[see][for
  details]{assef10}. Figure \ref{fg:ahat_r90_c90} shows the
distribution of $\hat{a}$ for our full-depth $R_{90}$ and $C_{90}$ AGN
candidate samples. We find that the $R_{90}$ sample is skewed towards
objects dominated by their AGN component, with almost no objects being
best-fit as inactive galaxies. This feature is also observed, although
to a somewhat lesser degree, in the $R_{75}$ sample.  The $C_{90}$
sample, on the other hand, shows a very considerable peak at
$\hat{a}=0$, as expected given its low reliability but high
completeness. It also shows, however, a very significant increase in
the number of objects with intermediate $\hat{a}$ values. These
dominate the distribution for $\hat{a}>0$. Most such objects probably
correspond to real AGN with high host fractions, implying that our
reliability optimized criteria is strongly biased against such
objects.

It is well known that the luminosity of the spheroidal component of
the host galaxy is correlated with the mass of its central SMBH, and that
this relation is roughly linear: $L_{\rm host} \sim M_{\rm BH}$
\citep[see, e.g.,][although also see \citealt{graham12} for possible
  deviations]{magorrian98,ferrarese05,graham07}. Some authors have
postulated that the correlation is also present, and non-evolving,
when considering the total host galaxy luminosity instead of just the
spheroidal component \citep{bennert10}. Regardless, since the
Eddington luminosity $L_{\rm Edd}$ is directly proportional to $M_{\rm
  BH}$, the Eddington ratio $\ell_{\rm Edd}$ can be expressed as
\begin{equation}\label{eq:eddington_ratio}
  \ell_{\rm Edd}\ =\ \frac{L_{\rm AGN}}{L_{\rm
      Edd}}\ \sim\ \frac{L_{\rm AGN}}{L_{\rm
      Host}}\ =\ \frac{\hat{a}}{1-\hat{a}}.
\end{equation}
\noindent Hence, to first order, AGN whose bolometric output is
dominated by the AGN emission (i.e., have high $\hat{a}$ values) also
correspond to objects emitting at a high $\ell_{\rm Edd}$. Similarly,
those galaxies for which stellar light represents a higher fraction of
their total bolometric output (i.e. low $\hat{a}$) are likely
radiating at lower Eddington ratios. So, in a physical context, we see
that our reliability-optimized selection criterion is strongly biased
against AGN radiating at low $\ell_{\rm Edd}$, but as our selection
criterion is shifted to emphasize high completeness, we start
recovering them.

Figure \ref{fg:ahat_r90_c90} also shows the distribution of $\hat{a}$
limited to $\rm W2<15.73$. While the $R_{90}$ sample looks nearly the
same, the $C_{90}$ sample exhibits a different distribution, with the
contamination ($\hat{a}=0$ peak) and the skewness of the distribution
shifting to be more similar to the $R_{90}$ sample. Partly, this is
because the largest contamination in completeness-optimized samples
comes from high-redshift galaxies, which are avoided by the $R_{90}$
criterion. That the peak at high $\hat{a}$ is increased is possibly
due to low Eddington ratio AGN simply being less luminous on average.

\begin{figure}
  \begin{center}
    \plotone{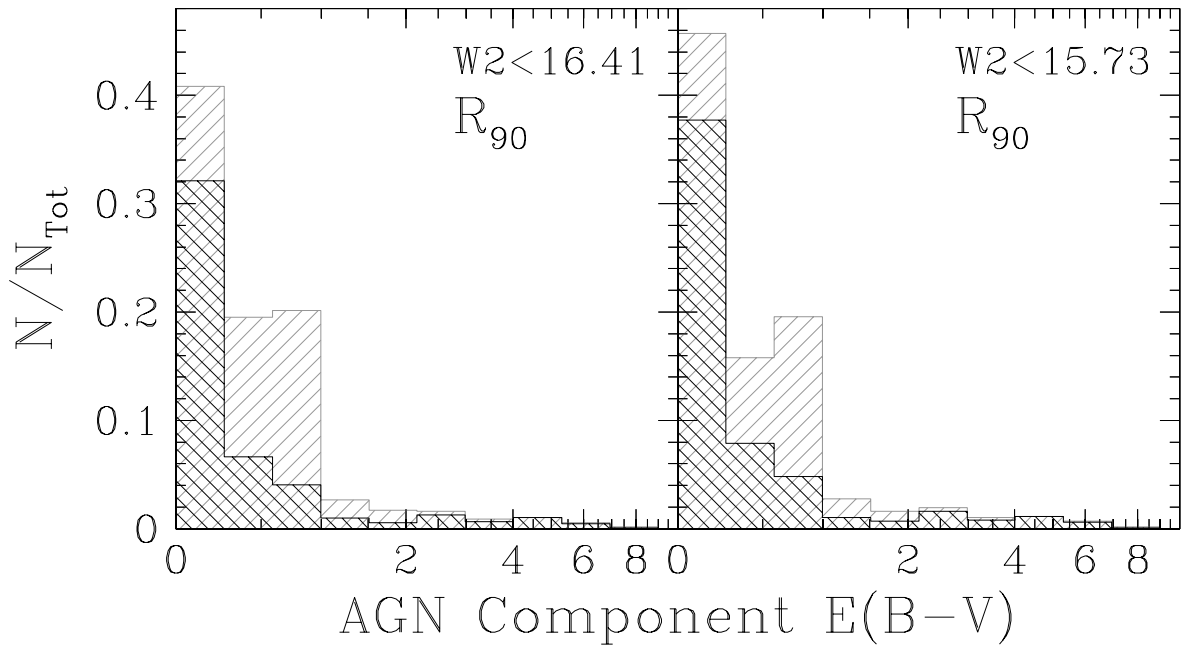}
    \caption{Distribution of best-fit AGN reddening $E(B-V)$ values
      for our $R_{90}$ AGN candidate samples. Left panel shows the
      full W2 depth sample, while the right one is limited to objects
      with $\rm W2<15.73$. Black histograms only include objects with
      spectroscopic redshifts, while gray histograms also include
      objects with photometric redshifts.}
    \label{fg:red_r90}
  \end{center}
\end{figure}

The other SED-fit parameter of interest is the amount of obscuration
towards the AGN, which is shown in Figure \ref{fg:red_r90} for the
$R_{90}$ AGN sample at full and 10$\sigma$ W2 depths. The most
important result to notice is that the WISE AGN selection is sensitive
even to objects with high obscuration. In order to interpret the
distribution, however, we need to deal with two issues. The first is
that because of the algorithm design, the reddening may be slightly
underestimated. Second, we need to take into account the selection
function of AGES, since the reddening obtained from objects with only
photometric redshifts estimates can be inaccurate
(\S\ref{ssec:photoz_acc}). We deal with both issues and present a
detailed study of the reddening properties of AGN in Section
\ref{sec:dust_properties}.

\subsection{Keck Observations}\label{ssec:keck}

To highlight the power of WISE in finding highly obscured quasars, we
obtained additional spectroscopy of 12 AGN candidates at the Keck
Observatory in April 2011.  Since the AGES spectroscopy is limited to
$I<22.5$, and is highly complete for $I<21.5$, we emphasized optically
fainter candidates which are bright in W2, selected on the basis of an
early version of the \citetalias{stern12} criteria. We furthermore
required them to not have a measured redshift. Because these
observations used a preliminary version of the WISE data, some of the
WISE colors and positions changed relative to the more accurate
all-sky release.  Therefore, the sources we observed have a range of
WISE colors and optical magnitudes; based on the WISE all-sky data
release, all but one of them are sufficiently red in W1$-$W2 to be
classified as AGN candidates by at least the $C_{90}$ criterion, but
not all of them are optically faint ($I>21$).  However, most of the
targets do meet the $R_{90}$ selection criterion and prove to be {\em
  bona fide} obscured quasars (see Table \ref{tab:deimos} for W1--W2
color, W2 magnitude and AGN classification criteria met by each
target).

We observed three Keck slitmasks in the Bo\"otes field with the DEep
Imaging Multi-Object Spectrograph \citep[DEIMOS;][]{faber03} on UT
2011 April 1-3.  We used the 4000 \AA\ order-blocking filter and the
600 $\ell$ mm$^{-1}$ grating (blazed at 7500 \AA; resolving power $R
\equiv \lambda / \Delta \lambda \sim 1600$ for the 1\farcs2 wide
slitlets we employed).  We observed a single additional mask using the
dual-beam Low Resolution Imaging Spectrometer \citep[LRIS;][]{oke95}
on UT 2011 April 28.  The LRIS observations employed the 300 $\ell$
mm$^{-1}$ grism on the blue arm of the spectrograph (blazed at 5000
\AA; $R \sim 500$), the 400 $\ell$ mm$^{-1}$ grating on the red arm of
the spectrograph (blazed at 8500 \AA; $R \sim 700$), and 6800
\AA\ dichroic.  Data reduction followed standard procedures, and we
flux calibrated the data using standard stars from \citet{massey90}.

Table \ref{tab:deimos} summarizes the results for these observations,
including measured redshifts, selection criteria, and the best-fit
$\hat{a}$ and AGN reddening parameters for the adopted redshift, with
errors derived from Monte Carlo re-sampling of the data. Appendix B
presents the results for additional Bo\"otes targets observed on these
masks.  We include the quality (``Q'') of each spectroscopic redshift.
Quality flag ``A'' signifies an unambiguous redshift determination,
typically relying upon multiple emission or absorption features.
Quality flag ``B'' signifies a less certain redshift determination,
such as the robust detection of an isolated emission line, but where
the identification of the line is uncertain
\citep[e.g.,][]{stern00}. Quality flag ``B'' might also be assigned to
a source with a robust redshift identification, but where some
uncertainty remains as to the astrometric identity of that
spectroscopic source. We consider the quality ``B'' results likely to
be correct, but additional spectroscopy would be beneficial. We assign
a quality flag ``F'' to all cases where a spectrocopic redshift could
not be reliably determined.

\begin{figure}
  \begin{center}
    \plotone{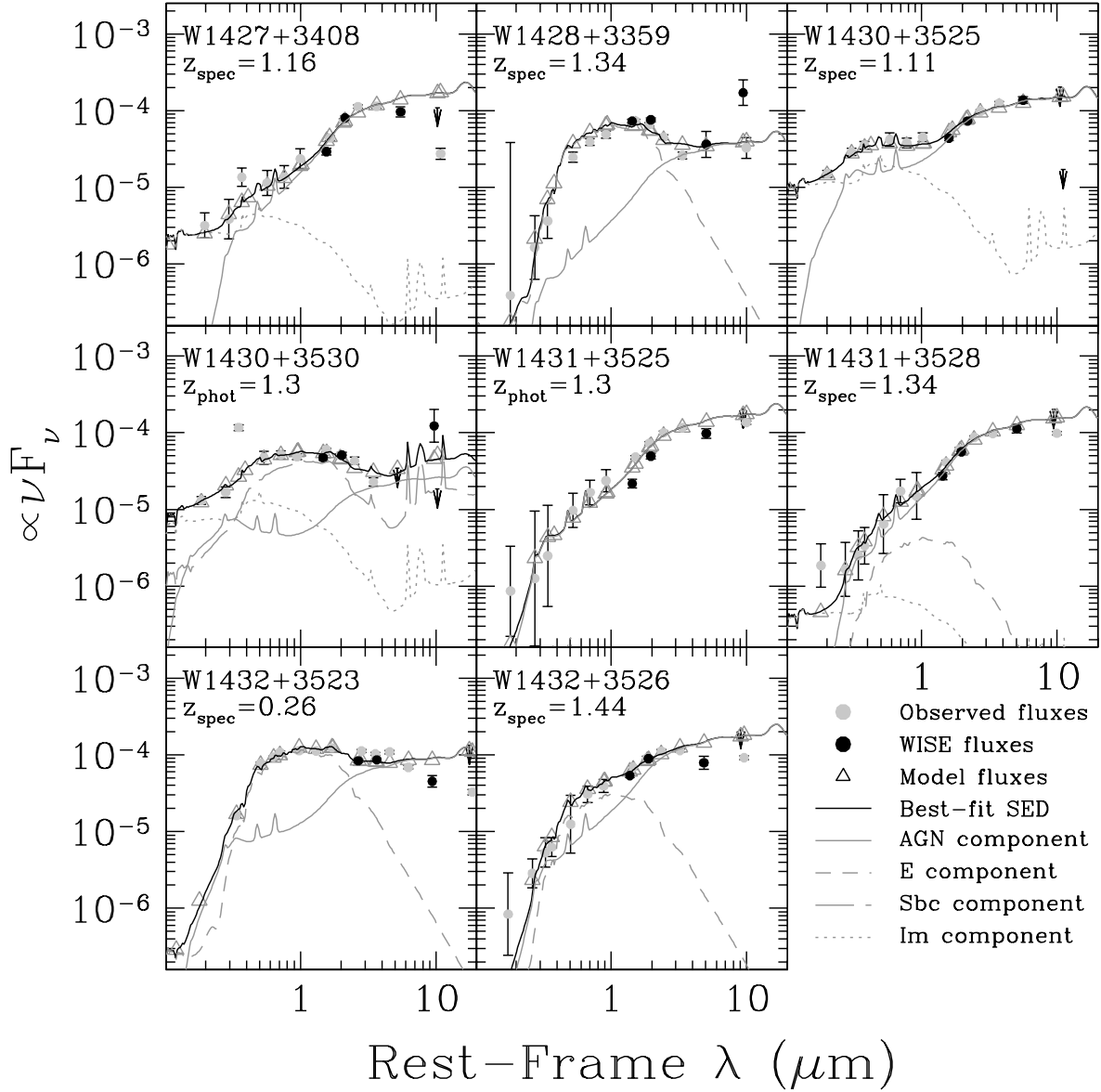}
    \caption{Best-fit SEDs for our sample of highly obscured,
      high-redshift AGN candidates observed with Keck. Only candidates
      that met one of the selection criteria beyond the very inclusive
      $C_{90}$ when using the all-sky data release WISE photometry are
      shown.}
    \label{fg:deimos_seds}
  \end{center}
\end{figure}

Figure \ref{fg:deimos_seds} shows the best-fitted SEDs for each of the
8 targets, from the original 12, whose all-sky release WISE W1$-$W2
colors classify them as AGN by either the $R_{90}$ or $R_{75}$
criterion. Upon inspecting the optical images, we believe the bright,
discrepant $I$-band flux of W1430+3530 is most likely due to a bright
star within 30\arcsec\ contaminating the photometry. Although most of
these objects appear to be real AGN based on their broad-band SEDs,
many lack strong, high-ionization lines such as C\,{\sc iv}, Mg\,{\sc
  ii} and Ne\,{\sc v}, even though lower ionization lines common for
star-formation are indeed observed (see Table \ref{tab:deimos}).  The
X-ray community has noticed a related population of X-ray bright,
optically normal galaxies \citep[XBONGs; e.g.,][]{civano07} where the
X-ray luminosities require the presence of an actively accreting SMBH
while optical spectroscopy reveals an apparently normal, inactive
galaxy.  Several explanations have been offered to explain such
sources, ranging from systematic effects that dilute the AGN signature
for the wide slit widths typically used for these distant sources
\citep[e.g.,][]{moran02}, to radiatively inefficient accretion flows
\citep[e.g.,][]{trump11}. Alternatively, at least some of these
objects could be better described as AGN-dominated LIRGs or ULIRGs,
where the lack of high ionization emission lines and the red host
color may be explained by large scale obscuration. Some evidence of
the Si 9.7$\mu$m absorption feature typical of ULIRGs may be present
in a few cases (W1427+3408, W1431+3525, W1432+3523 and W1432+3526),
causing discrepancies between the models and the data, although this
feature may also be observed in AGN under certain conditions
\citep[see, e.g.,][]{feltre12}. Some of the discrepancies observed,
however, such as W4 and MIPS 24$\mu$m for W1428+3359, W1430+3525 and
W1432+3523, and W3 for W1432+3526, are possibly simply due to the
inherent difficulties of mid-IR observations.

\section{Dust Reddening in AGN}\label{sec:dust_properties}

In this section we study dust obscuration properties of a set of 362
$z<1$ AGN well detected by WISE with spectroscopic redshifts, $I<20$
and $\hat{a}>0.5$ in the Bo\"otes field. As argued earlier, low
redshift ($z\lesssim 1$) WISE AGN selection criteria are relatively
insensitive to obscuration since they rely on the hot dust emission
from the dust torus instead of on the blue colors of the unobscured
accretion disk emission, as per optical selection. Hence, we can use
WISE to study the properties of dust obscuration in AGN.

AGN unification models \citep[see, e.g.,][]{antonucci93,urry95}
propose that Type 1 and Type 2 AGN are physically equal but are
observed at different inclination angles relative to the obscuring
material near the AGN. Typically it is assumed that the accretion
disk, responsible for the $\lambda \lesssim 1\mu\rm m$ continuum
emission, extends to radial scales of $\sim 10~\rm AU$, and is
surrounded by highly ionized gas responsible for the broad
emission-lines. On larger scales ($\sim 1~\rm pc$) there is dust in
the general shape of a ``torus'' or a flared disk, responsible for the
Type 1/Type 2 dichotomy, that absorbs the optical radiation from the
accretion disk and re-emits it in the mid-IR. We refer to this
structure as the torus, as is commonly done, although we do not {\it{a
    priori}} assume a shape for it. The inner edge of the dust torus
is determined by where the dust reaches its sublimation temperature
due to heating from the accretion disk. Such hot dust produces the
emission observed to dominate the mid-IR portion of the AGN
SED. Furthermore, narrow emission lines are observed in both Type 1
and Type 2 AGN, and are known not to be polarized
\cite[e.g.,][]{antonucci93}, so the dust structure must be smaller
than the narrow line region ($\sim 1~\rm kpc$). Given that the Type
1/Type 2 dichotomy is also manifested in the neutral hydrogen
absorption of the X-ray emission, the torus must also be associated
with the absorbing gas.

Many properties of the dust torus have been extensively studied. For
example, several authors
\citep[e.g.,][]{krolik88,nenkova02,nenkova08,elitzur06,tristram07}
have argued that the dust in the torus must be in optically and
geometrically thick clumps to reproduce observations, while others
\citep[e.g.,][]{dullemond05,fritz06} argue the dust may be smoothly
distributed. A recent study by \citet{feltre12}, however, suggests
that given the same dust composition and the same illuminating source,
the difference in the broad-band shape of the SEDs from these dust
configurations may be too subtle to distinguish between scenarios with
current data. In a more global sense, the geometry and evolution of
the obscuring structures have also been studied, as, for example, the
fraction of obscured objects can have profound implications for
explaining the cosmic hard X-ray background \citep[see,
  e.g,][]{ueda03}. \citet{simpson05} has shown using the Type 1 and
Type 2 AGN optical luminosity function from SDSS that the fraction of
Type 2 AGN increases with decreasing accretion disk luminosity, and a
similar behavior has been observed for radio galaxies
\citep{lawrence91,simpson98,grimes04} and in the X-rays
\citep[e.g.,][]{ueda03,hasinger04}. Such a behavior can be naturally
expected if the scale height $h$ of the obscuring material is
independent (or not linearly related) to the radial size of the
structure ($R\propto \sqrt{L_{\rm AGN}}$), such that for brighter AGN
the dust effectively covers a smaller solid angle as viewed from the
SMBH. This scenario is usually referred to as the ``receding torus
model'', and was first proposed by \citet{lawrence91}. In particular,
\citet{simpson05} has shown that observations appear to be best
reproduced if $h\propto L_{AGN}^{0.23}$.

Combining WISE and all the ancillary observations in the NDWFS
Bo\"otes field and performing the SED modeling as detailed in
\S\ref{ssec:sed_fits}, we can study the average properties of dust
obscuration in AGN by counting the number of objects observed per unit
reddening. This approach allows us to quantify the fraction of AGN
that can be classified as Type 1, and also, in principle, to
differentiate between different dust geometries and compositions. Note
that obscuring dust may also be present in the interstellar medium
(ISM) of the respective host galaxies, and while we expect AGN
obscuration to be mainly driven by the dust in the torus and hence
delineate the discussion in that direction, we further address
galactic-scale obscuration in \S\ref{ssec:red_results}. Since our
sample is, in essence, flux limited, our analysis must properly
take into account its selection function. Specifically, it must
account for all the biases against highly obscured objects, since
higher AGN obscuration can also make the objects appear much fainter
depending on the relative contribution and SED shape of the host
galaxy. Fortunately, the SED fitting approach of \citet[][see also
  \S\ref{ssec:sed_fits}]{assef10} is well suited to assess and correct
for our survey incompleteness. In the next section we detail our
sample selection function. In \S\ref{ssec:red_method} we detail the
formalism we use to incorporate the selection function in our
measurement of the reddening distribution, while in
\S\ref{ssec:red_results} we show and discuss the resulting
distributions.

\subsection{Sample Selection Function}\label{ssec:red_sample}

To study the reddening distribution, we use a subsample of the larger
sample described in \S\ref{sec:data}. We require that objects have
$\rm W2<15.73$, a measured spectroscopic redshift such that $E(B-V)$
is accurately estimated (see \S\ref{ssec:photoz_acc}), and
$\hat{a}>0.5$ to minimize possible non-AGN contaminants. Note that we
do the initial selection with the SED fits obtained including all the
priors described in \S\ref{ssec:sed_fits}, which is necessary to
ensure all obscured AGN are real. As discussed there, this can lead to
slightly underestimated AGN obscuration. Hence, once the sample is
selected, we re-fit the SEDs removing all priors described in
\S\ref{ssec:red_results} to obtain the final $E(B-V)$ values, although
not removing the prior does not qualitatively affect our
results. Note, however, that this will cause some incompleteness at
the highest obscuration ($E(B-V)\gtrsim 5$) end of our sample. We
visually inspected the SED fit of every source and eliminated 16
galaxies where we believed the AGN classification or the reddening
values were spurious due to bad photometry.

We further require that the redshift was determined by AGES, since its
well determined selection function is a crucial component of our
analysis. Since many of the objects we consider are extended in the
NDWFS imaging and were not necessarily targeted as AGN candidates by
the AGES survey, we must also restrict our sample to objects with
$I<20$, resulting in a final sample size of 362 AGN. AGES was designed
to ensure subsamples are statistically complete to $I<20$ for galaxies
and $I<21.5$ for AGN \citep[see][for details]{kochanek12}. In order to
do this, AGES used a sparse sampling algorithm for galaxies, such that
for every defined galaxy subsample, a spectrum was attempted for all
objects brighter than a certain magnitude limit and for a percentage
(typically 20--30\%) of randomly selected fainter galaxies down to a
certain magnitude. For example, the main $I$-band selected galaxy
sample was observed in full for $I<18.5$ and 20\% of the galaxies were
followed in the range $18.5<I<20$. In contrast, there was no sparse
sampling for AGN candidates as AGES attempted to get spectra of all of
them. Every subsample was assigned a selection code, where $P_{i, \rm
  sparse}^{n}$ is the probability that object $i$ of the subsample
with selection code $n$ was selected for spectroscopy due to the
sparse sampling algorithm.

In addition, the fraction of sources with a successfully measured
redshift depends on $I$-band magnitude. While the survey design
minimized the magnitude dependence beyond the sparse sampling, there
is still a dependency simply because it is more difficult to obtain
redshifts for fainter sources in a fixed integration time. Using the
full results of the AGES survey, we estimate for every selection code
the fraction of objects for which spectra were attempted and a
redshift was measured as a function of $I$-band magnitude,
$P_{i,I}^{n}$.

In order to correct for the selection function, we need to estimate
for every object the probability that objects with the same optical
and IR magnitudes would have been observed, so that we can
statistically account for those without spectroscopic
observations. Since every object may have been targeted for more than
one of the different subsamples, we need to consider the joint
probability of all subsamples. Let $P_{i}^{n} = P_{i, \rm sparse}^{n}
\times P_{i, I}^{n}$ and let $N$ be the total number of subsamples
object $i$ is part of. We define $C(N,k)$ to be the sum of all
possible combinations of $k$ element products of the $P_{i}^{n}$
terms, such that, for example, $C(3,1) = P_{i}^1 + P_{i}^2 + P_{i}^3$,
$C(3,2) = P_{i}^1 P_{i}^2 + P_{i}^1 P_{i}^3 + P_{i}^2 P_{i}^3$ and so
on. The probability a spectroscopic redshift would have been obtained
for objects like object $i$ is then given by
\begin{equation}\label{eq:p_i}
  P_{i}\ =\ \sum_{k=1}^{N}\ (-1)^{k+1}\ C(N,k).
\end{equation}
\noindent It can be shown that if any of the terms $P_{i}^{n}=1$, then
$P_{i}=1$, as would be expected. For consistency with the original
AGES selection, we use the original catalogs of AGES to assess the
spectroscopic completeness rather than the catalogs described in
\S\ref{sec:data}.

\subsection{Method}\label{ssec:red_method}

In order to incorporate the selection function, we adapt the step-wise
maximum likelihood method (SWML) of \citet{efstathiou88}. For the
remainder of this section, we define $E_{BV}\equiv E(B-V)$. Our goal
is to estimate the distribution $\xi(E_{BV}) = dn/dE_{BV}$, where we
remind the reader that $E_{BV}$ corresponds to the reddening only over
the AGN component, not over the host galaxy (see \S\ref{ssec:sed_fits}
for details). The probability of finding an object with a given
reddening $E_{BV}^i$ is given by
\begin{equation}\label{eq:prob_i}
  p_i\ \propto\ \left(\frac{\xi(E_{BV}^i)}{\int_0^{E_{BV, \rm Max}^i}
    \xi(E_{BV}) dE_{BV}}\right)^{C_i},
\end{equation}
\noindent where $C_i = P_i^{-1}$ is calculated using equation
(\ref{eq:p_i}) and $E_{BV, \rm Max}^{i}$ is the maximum reddening
object $i$ could have and still be in our sample, which we detail
below. We estimate $E_{BV, \rm Max}$ by varying the reddening of the
AGN component of the best-fit combination of SED templates but keeping
the amplitude of the components fixed (see \S\ref{ssec:sed_fits} for
details).

To apply the SWML method we discretize the function $\xi(E_{BV})$ in
bins of $E_{BV}$ rather than assuming a parametric form. We divide
$\xi(E_{BV})$ into $N_p$ bins of value $\xi_k$, centered at reddening
values $\epsilon_k$ with widths $\Delta E_{BV}$. We can then rewrite
equation (\ref{eq:prob_i}) as
\begin{equation}
  p_i\ \propto\ \left(\frac{\sum_{k=1}^{N_p} W(E_{BV}^i -
    \epsilon_k)\ \xi_k}{\sum_{j=1}^{N_p}\ H(\epsilon_j - E_{BV, \rm
      Max}^{i})\ \xi_j\ \Delta E_{BV}}\right)^{C_i},
\end{equation}
\noindent where
\begin{equation}
  W(x)\ =\ \left\{ 
  \begin{array}{rl}
    1 & {\rm{if}}\ -\Delta E_{BV}/2 \leq x \leq \Delta E_{BV}/2,\\ 0 &
    {\rm{otherwise}}
  \end{array}\right.
\end{equation}
\noindent and
\begin{equation}
  H(x)\ =\ \left\{ 
  \begin{array}{rl}
    1 & {\rm{if}}\ x < -\Delta E_{BV}/2,\\ \frac{1}{2} -
    \frac{x}{\Delta E_{BV}} & {\rm{if}}\ -\Delta E_{BV}/2 \leq x \leq
    \Delta E_{BV}/2,\\ 0 & {\rm{if}}\ x > \Delta E_{BV}/2.
  \end{array}\right.
\end{equation}
\noindent The likelihood ${\mathcal{L}}$ of our sample being drawn
from the distribution $\xi(E_{BV})$ corresponds to the multiplication
of the $p_{i}$ values of all $N_{\rm AGN}$ objects in our
sample. Taking the gradient of ${\mathcal{L}}$ with respect to
$\vec{\xi} = (\xi_1,\dots,\xi_k)$, we can find that the values that
maximize the likelihood are given by
\begin{equation}
  \xi_k \Delta E_{BV}\ =\ \frac{\sum_{i=1}^{N_{\rm
        AGN}}\ C_i\ W(E_{BV}^i - \epsilon_k)}{\sum_{i=1}^{N_{\rm
        AGN}}\ \frac{C_i\ H(\epsilon_k - E_{BV, \rm
        Max}^i)}{\sum_{j=1}^{N_p}\ H(\epsilon_j - E_{BV, \rm
        Max}^{i})\ \xi_j\ \Delta E_{BV}}}.
\end{equation}
\noindent As discussed by \citet{efstathiou88}, a constraint is needed
since the likelihood only depends on the ratios of the $\xi_k$
values. We adopt the constraint
\begin{equation}
  g(\ \vec{\xi}\ )\ =\ \sum_{k=1}^{N_p} \xi_k \ -\ N_{\rm AGN},
\end{equation}
\noindent so that the sum of the bins simply equals the number of AGN,
and we maximize
$\ln{\mathcal{L}^{\prime}}\ =\ \ln{\mathcal{L}}\ +\ \lambda
g(\ \vec{\xi}\ )$, where $\lambda$ is a Lagrange multiplier. Errors in
$\xi_k$ are estimated using the information matrix as detailed in
\citet{efstathiou88}. In practice, since most of our objects have
relatively low reddening values, we prefer to estimate
$\xi^{\prime}(E_{BV}) \equiv dn / d \log (E_{BV} + 0.1)$. Finally, as
discussed by \citet{assef11b}, our AGN SED template is as blue as
possible, so some of the reddening we find is possibly just due to
intrinsic differences in the SEDs of Type 1 AGN. For example, the mean
Type 1 SED template of \citet{richards06} is similar to our AGN
template with $E(B-V)\approx 0.05$. Since we do not want this to bias
the results, we subtract 0.05 from all the $E(B-V)$ values before we
construct $\vec{\xi}$.

\subsection{Results}\label{ssec:red_results}

\begin{figure}
  \begin{center}
    \plotone{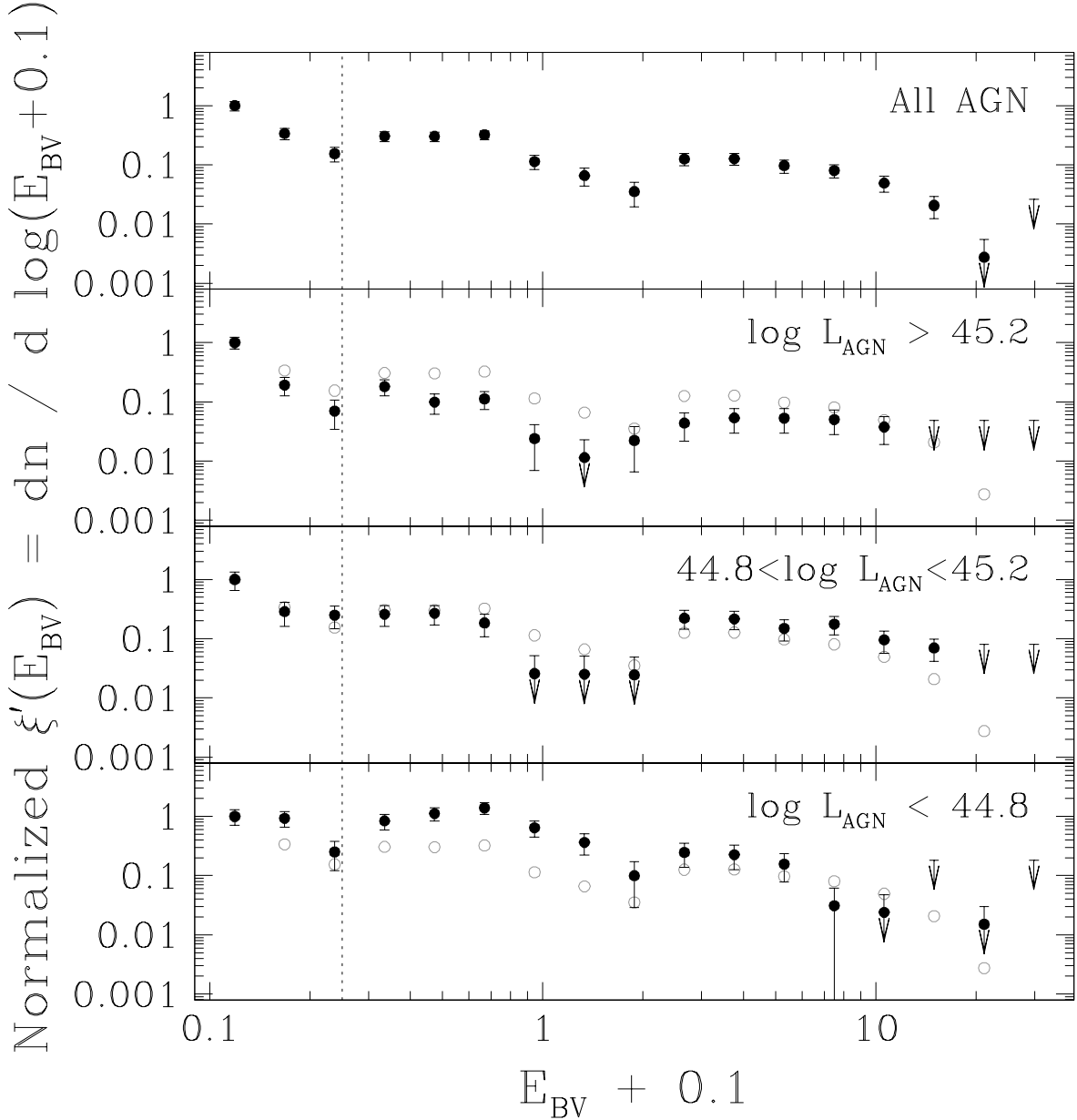}
    \caption{Distribution of reddening values ({\it{solid black
          circles}}) obtained after using the non-parametric formalism
      described in \S\ref{ssec:red_method} to account for sample
      incompleteness. For simplicity, we adopt the notation
      $E_{BV}\equiv E(B-V)$. The top panel shows the distribution
      obtained when using the complete AGN sample, while the lower
      three panels show the distributions in three bins of AGN
      bolometric luminosity (erg~s$^{-1}$) as defined in
      \S\ref{ssec:sed_fits}. The distribution obtained using all
      objects is repeated as open gray circles in each of the lower
      three panels for comparison. All the distributions are
      normalized to unity in the lowest $E_{BV}$ bin. The vertical
      dotted line shows our adopted reddening boundary between Type 1
      and Type 2 AGN.}
    \label{fg:dnde}
  \end{center}
\end{figure}

Figure \ref{fg:dnde} shows the distribution of $\xi^{\prime}(E_{BV})$
derived using the sample described in \S\ref{ssec:red_sample}. The key
thing to note is that the distribution falls with increasing
reddening, with a dip at $E(B-V)\sim 2$. A smaller, less significant
dip may also be present at $E(B-V)\sim 0.15$. Note that there are no
objects observed with a best-fit $E(B-V)>14$. Luminous objects with
such high reddening are expected to be rare \citep[see, e.g.,][for
  such extreme cases]{eisenhardt12,jingwen12,bridge13}. While the
general trend of decreasing numbers with increasing reddening is in
all likeliness real, our sample is small enough that the observed dips
could in principle be systematic and caused by the non-parametric
method we used, as it never imposes the requirement of a smooth
distribution. However, when the sample is divided into three
luminosity bins with equal numbers of objects, as shown also in Figure
\ref{fg:dnde}, the minimum of the dust distribution at $E(B-V)\sim 2$
appears in all of them, further suggesting this feature is
real. Assuming that is the case, a few possible explanations are
possible.

In the simplest orientation models for AGN unification, most of the
obscuration comes from the dust torus. However, if the minimum at
intermediate $E(B-V)$ is real, it is unlikely that the dust forms a
continuous medium, as it is very hard to have a physically motivated
dust distribution that produces such a feature. If the dust is, on the
other hand, in geometrically and optically thick clouds, the
distribution would simply be the distribution of the obscuration of
the clouds convolved with the distribution of inclination angles and
covering fractions. However, this is also unlikely to be consistent
with a minimum in the distribution at an intermediate obscuration
value. Possibly, thick dust clouds are responsible for the
$E(B-V)\gtrsim 2$ obscuration, and these are embedded in a diffuse
inter-cloud dust medium which is responsible for the lower obscuration
part of the $\xi^{\prime}(E_{BV})$ distribution.

Alternatively, the two halves of the distribution could be attributed
to different sources, with the large obscuration coming from thick
dust clouds in a torus-like structure surrounding the AGN, and the
lower obscuration coming from diffuse dust in the host galaxy
ISM. Naively, one would not expect the distribution of ISM dust
obscuration to vary systematically with AGN luminosity. When we divide
the sample in three bins of luminosity with equal numbers of objects,
as shown in Figure \ref{fg:dnde}, we observe a significantly different
shape for the $\xi^{\prime}(E_{BV})$ distribution in each bin. We
consider this as evidence that the dust obscuration is primarily
coming from the vicinity of the AGN and is hence associated with the
torus, and we discuss this below in the context of a receding
torus. It may be possible that in certain AGN feedback scenarios the
column density of the residual dust in the ISM left after the AGN has
gone through the blow-out phase \citep[see, e.g.,][]{hopkins08} could
be related to AGN luminosity during its quasar phase.

Note that since our sample is inherently magnitude limited, we cannot
easily disentangle redshift evolution from luminosity evolution. We
consider, however, that it is much more likely that the evolution in
the dust obscuration is primarily driven by the AGN luminosity since
hardly any evolution is observed in the UV through mid-IR SEDs of AGN
with cosmic time \citep[e.g.,][]{richards06,assef10}. Furthermore,
\citet{ueda03} has shown that the distribution of neutral gas column
densities obscuring the X-ray emission of AGN is independent of
redshift.

We investigate the fraction of Type 1 to Type 2 AGN by simply adding
up the corresponding bins of the $\xi^{\prime}(E_{BV})$
distribution. We adopt the standard X-ray boundary of a gas column
density of $N_H = 10^{22}~\rm cm^{-2}$ \citep[e.g.,][]{ueda03} as the
dividing line. \citet{maiolino01} has shown that the value of
$E(B-V)/N_H$ is significantly below the Galactic value for AGN, and
varies significantly among different AGN. The median value of the
\citet{maiolino01} sample is $E(B-V)/N_H = 1.5\times 10^{-23}~\rm
cm^2~\rm mag$, which puts the Type 1/2 boundary at $E(B-V) = 0.15$, or
$A_V = 0.47$ for $R_V=3.1$. It also implies that our sample does not
contain any Compton-thick AGN ($N_H>10^{24}~\rm cm^{-2}$), which is a
reasonable expectation given the requirement of $\hat{a}>0.5$ and the
bias of our method to underestimate this value for the most highly
obscured AGN (see \S\ref{ssec:sed_fits} and Appendix A). From the
joint distribution of all AGN, we find that the fraction of objects
that would appear as Type 1 AGN is $47\pm 8$\%, consistent with an
even split between the two types. From a purely observational point of
view, this is only strictly appropriate for $I<20$ AGN given our
sample selection. However, we note that very little variation in this
ratio with $I-$band magnitude is observed in our sample.

\begin{figure}
  \begin{center}
    \plotone{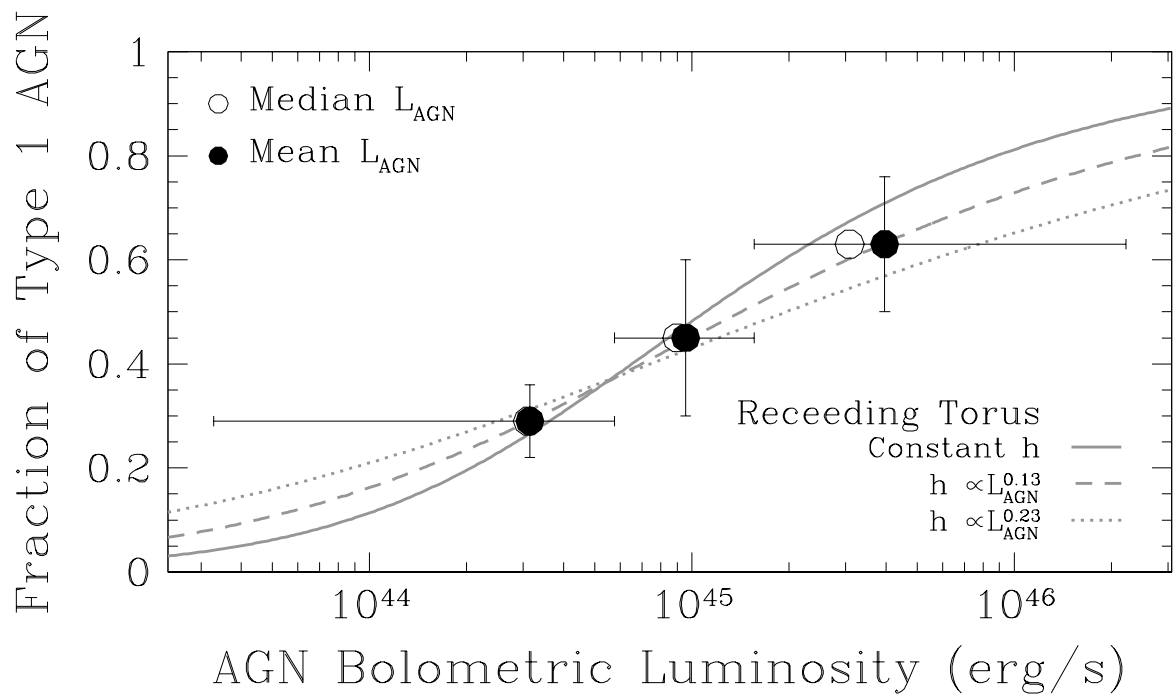}
    \caption{Type 1 AGN fraction as a function of AGN bolometric
      luminosity, as defined in \S\ref{ssec:sed_fits}. The solid
      points are centered at the mean AGN bolometric luminosity of the
      bin while the open ones are centered at the median value. The
      luminosity error-bars show the range of each luminosity bin. The
      gray lines shows the best-fit receding torus models described in
      the text.}
    \label{fg:t1_frac}
  \end{center}
\end{figure}

Figure \ref{fg:t1_frac} shows the fraction of Type 1 AGN as a function
of luminosity when we divide the sample into three luminosity bins
with equal numbers of objects per bin. There is a sharp increase in
the Type 1 fraction towards higher luminosities. For the lowest
luminosity bin we find that the fraction of objects appearing as Type
1 is $29\pm 7$\%, increasing to $46\pm 15$\% for the intermediate
luminosity bin, and to $64\pm 13$\% at the highest luminosity. This
trend conforms to the idea of a receding torus. Figure
\ref{fg:t1_frac} compares our observed trends with three different
models of a receding torus, taken from \citet{simpson05}. We model the
fraction of Type 1 AGN by
\begin{equation}
  f_1\ =\ 1\ -\ \left[1\ +\ 3 \left(\frac{L_{\rm AGN}}{L_{\rm
        AGN,0}}\right)^{1-2\psi}\right]^{-0.5},
\end{equation}
\noindent which comes from the simple geometry assumed by
\citet{simpson98} and assuming the scale height $h\propto L_{\rm
  AGN}^{\psi}$. At luminosity $L_{\rm AGN,0}$, AGN are evenly split
between Types 1 and 2. We first consider the two cases studied by
\citet{simpson05}, namely that of a constant $h$ ($\psi=0$), and his
favored scenario of $\psi=0.23$. Note that because \citet{simpson05}
used [O\,{\sc iii}] luminosities as proxies for the accretion disk
luminosity, we must fit for $L_{\rm AGN,0}$, obtaining respectively
$1.12^{+0.16}_{-0.14}\times 10^{45}~\rm erg~\rm s^{-1}$ ($\psi=0$) and
$1.98^{+0.57}_{-0.44}\times 10^{45}~\rm erg~\rm s^{-1}$
($\psi=0.23$). As shown in Figure \ref{fg:t1_frac}, both of them give
a fair representation of the data. If we also fit for the dependence
of $h$ on $L_{\rm AGN}$, we find $\psi=0.13\pm 0.17$ and $L_{\rm
  AGN,0} = 1.42^{+10.9}_{-1.26}\times 10^{45}~\rm erg~\rm
s^{-1}$. Unfortunately, our modest sample size does not allow us to
more finely sample the Type 1 AGN fraction as a function of AGN
luminosity and thereby further constrain such models.  We do note,
however, that the reddening distributions shown in Figure
\ref{fg:dnde} have significant power to further constrain the dust
distribution.  This will be be further explored in future work.

It is worth noting that \citet{treister04} found that a non-evolving
Type 1 fraction of 25\% yielded consistency between the soft X-ray and
$z$-band flux distributions of AGN. Given that the AGN in that study,
performed in the GOODS fields, are typically of lower luminosity than
the AGN in our sample, this is in general agreement with the $29\pm
7\%$ we find for our lowest luminosity bin. In contrast,
\citet{hopkins07} found that an obscured fraction of
$0.26~(L/10^{46}~\rm erg~\rm s^{-1})^{0.082}$ at rest-frame
4000\AA\ brings luminosity functions at different wavelengths into
good agreement. This value is inconsistent with ours, although a
detailed comparison is hard to make as their values are also affected
by scatter and luminosity dependence of their assumed bolometric
corrections. Further comparison with theoretical expectations to match
the hard X-ray background would be useful, but are beyond our reach
given our insensitivity to Compton-thick AGN.

\section{Conclusions}

In an earlier study (\citetalias{stern12}) we used the extensive
spectroscopy and photometry of the 2 deg$^2$ COSMOS field to study
WISE AGN selection. We found that the simple criterion W1--W2$\geq$0.8
and W2$<$15.05 produces a sample with 95\% reliability and recovered
78\% of the AGN found with {\it{Spitzer}} IRAC imaging to the same
flux depth. Here we have extended this study using the larger 9
deg$^2$ NDWFS Bo\"otes field, which has also significantly deeper WISE
observations than COSMOS. We show that the reliability of a simple
color cut quickly degrades towards fainter fluxes due to the large
number of $z\gtrsim 1$ galaxies that contaminate the color selection.

Using the extensive UV through mid-IR broad-band photometry available
in the NDWFS Bo\"otes field we have studied W2--dependent W1--W2
selection criteria optimized to find AGN at deeper WISE fluxes than
those available in the COSMOS field. We provide different criteria
depending on whether the emphasis is on reliability or
completeness. We defined a reliability-optimized criteria as a W1--W2
color limit that varies as an exponential of $\rm W2^2$, where the
parameters can be tuned to achieve different reliability levels
(\S\ref{ssec:mag_dep_cut}). We find that for completeness-optimized
selection, no dependence on W2 is needed; a simple W1--W2 color
criterion suffices. We find that the criterion of \citetalias{stern12}
returns samples with a completeness of approximately 75\%.

We have also studied the accuracy of broad-band photometric redshifts
obtained for the WISE AGN candidates using the \citet{assef10} basis
of low-resolution SED templates for AGN and galaxies. We find
consistency with the poor accuracy found by previous studies, even
though our AGN are brighter than those typically used in such
studies. Furthermore, we find that although the value of the $\hat{a}$
parameter, the luminosity fraction of the AGN with respect to the host
plus the AGN, is insensitive to uncertainties in the photometric
redshift, the best-fit reddening of the AGN component is strongly
affected by those uncertainties. We have studied the distribution of
the best-fit $\hat{a}$ parameter of the WISE AGN candidates, showing
they are biased towards high values. This means that WISE AGN
selection is biased towards objects that are bright with respect to
their hosts. Since the luminosity of the host is roughly correlated
with the mass of its central SMBH \citep[e.g.,][]{magorrian98}, this
can probably be expressed as a bias towards AGN radiating at large
fractions of their Eddington limits.

Finally, we have studied the distribution of AGN reddening in the WISE
AGN candidates. We have shown that although WISE is more sensitive to
unobscured objects, it still finds considerable numbers of highly
obscured objects. Extending the sample to include all AGN found over
the field with $\hat{a}>0.5$, spectroscopic redshifts from the AGES
survey, and high $S/N$ WISE W2 fluxes, we have studied the
distribution of objects as a function of AGN reddening. We present a
formalism based on the step-wise maximum likelihood method of
\citet{efstathiou88} designed to account for sample incompleteness as
a function of obscuration. For a subsample of 362 objects with $I<20$
and W2$<$15.73 for which the selection function is well understood, we
find that the reddening distributions depend on AGN bolometric
luminosity. The distribution is peaked for unobscured objects and then
falls relatively monotonically towards $E(B-V)\sim 2$, raising towards
higher values and then dropping again towards $E(B-V)\sim 10$. While
it is possible that our small sample size could be driving some of the
observed structure, we point out that this shape could be explained by
continuous diffuse dust medium in which optically thick dust clouds
are embedded. We find that when looking at the complete subsample,
47$\pm$8\% of AGN are Type 1 ($E(B-V)<0.15$; see
\S\ref{ssec:red_results}). This fraction is a strong function of the
AGN bolometric luminosity, consistent with the general scenario of a
receding torus. At our lowest luminosity bin, centered at $L_{\rm AGN}
= 3\times 10^{44}~\rm erg~\rm s^{-1}$, we find a Type 1 fraction of
29$\pm$7\%, which rises to 64$\pm$13\% for the highest luminosity bin
centered at $L_{\rm AGN} = 4\times 10^{45}~\rm erg~\rm s^{-1}$. Larger
samples, such as that provided by the combination of SDSS and WISE,
will provide greater constraints and insight into the dust
distribution in AGN.

\acknowledgements We would like to thank M.~Dickinson, A.H.~Gonzalez,
J.~Kartaltepe, B.~Mobasher, H.~Nayyeri, K.~Penner and G. Zeimann for
helping us obtain some of the Keck spectroscopic observations used in
this work. We thank M.~Elitzur for an insightful discussion about dust
properties in AGN. We thank the NDWFS, NEWFIRM and MAGES survey teams
for providing their respective data sets over the Bo\"otes field. We
thank the anonymous referee for suggestions that helped improve our
work. RJA and CWT are supported by an appointment to the NASA
Postdoctoral Program at the Jet Propulsion Laboratory, administered by
Oak Ridge Associated Universities through a contract with NASA. This
publication makes use of data products from the Wide-field Infrared
Survey Explorer, which is a joint project of the University of
California, Los Angeles, and the Jet Propulsion Laboratory/California
Institute of Technology, funded by the National Aeronautics and Space
Administration. Some of the data presented herein were obtained at the
W.M. Keck Observatory, which is operated as a scientific partnership
among the California Institute of Technology, the University of
California and the National Aeronautics and Space Administration. The
Observatory was made possible by the generous financial support of the
W.M. Keck Foundation.

\appendix

\section{Accuracy of the AGN--Host-galaxy SED Decomposition}

\begin{figure}
  \begin{center}
    \plotone{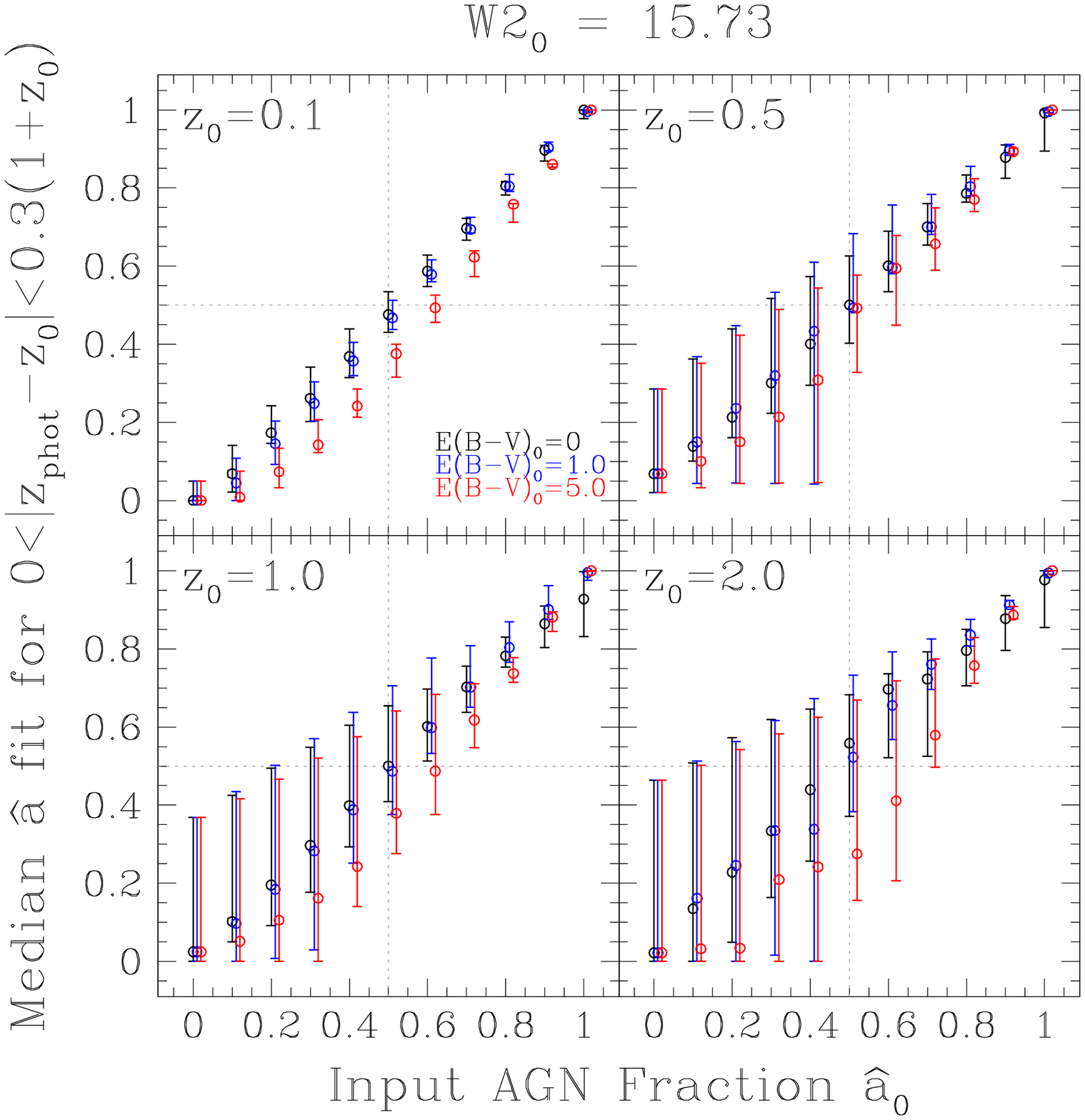}
    \figurenum{A1}
    \caption{Median $\hat{a}$ values obtained for photometric
      redshifts within $0.3~(1+z_0)$ of the intrinsic redshift $z_0$
      for a set of simulated galaxies with a given $\hat{a}_0$ AGN
      fraction and $E(B-V)_0$ AGN obscuration. We have assigned a
      W2$_0$ magnitude of 15.73 for all simulated objects. The
      error-bars show the range encompasing 95.4\% of the trials.}
    \label{fg:ahat_acc_sn10}
  \end{center}
\end{figure}

\begin{figure}
  \begin{center}
    \plotone{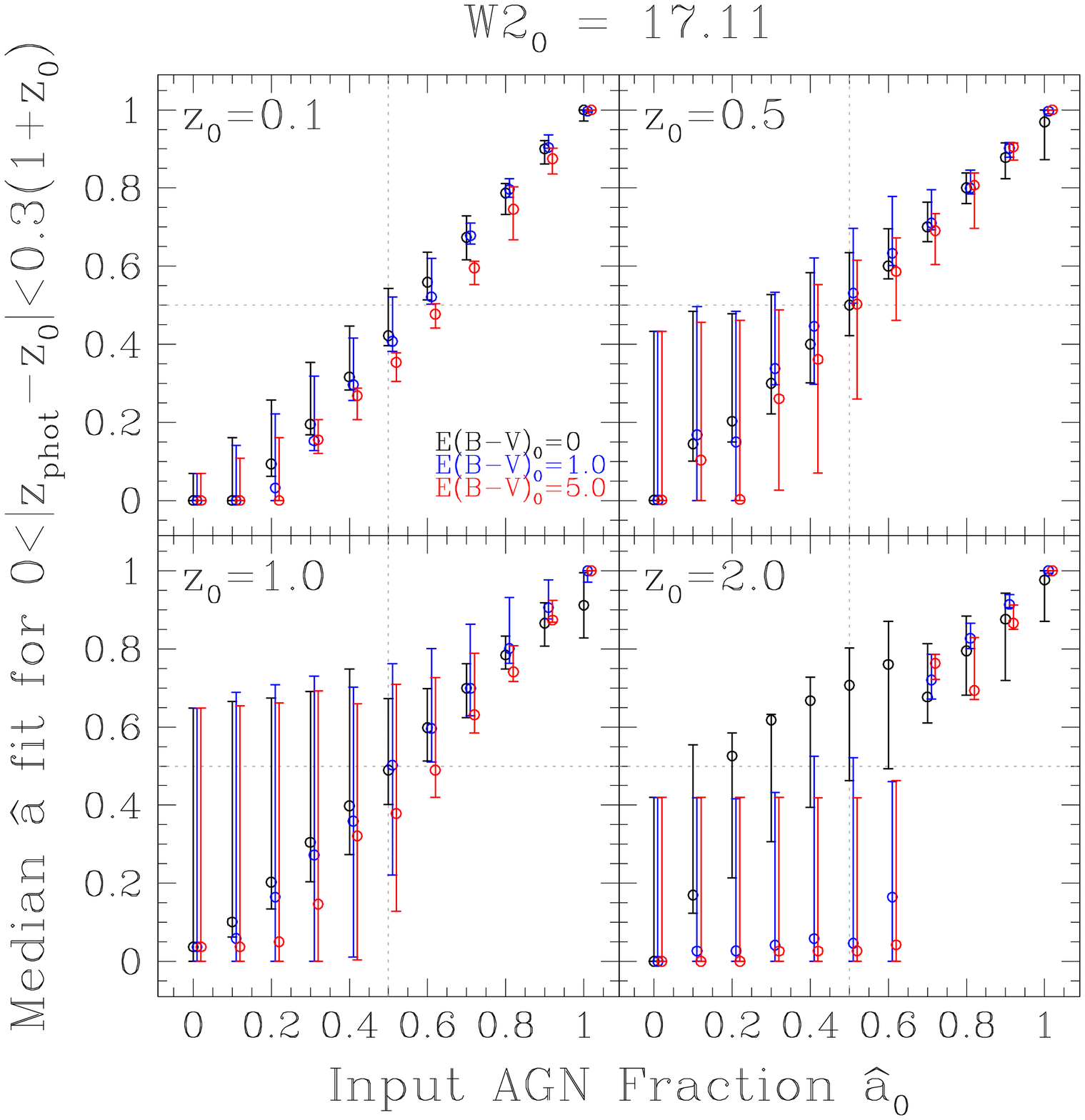}
    \figurenum{A2}
    \caption{Same af Fig. \ref{fg:ahat_acc_sn10} but for simulated
      objects with an assigned W2$_0$ magnitude of 17.11.}
    \label{fg:ahat_acc_sn3}
  \end{center}
\end{figure}

A proper characterization of accuracy of the $\hat{a}$ determination
in the presence of photometric redshift errors is difficult to
quantify beyond the work already presented in \citet{assef10}, as it
depends on several different factors, such as AGN obscuration,
``true'' redshift of the source and the intrinsic value of
$\hat{a}$. However, we can do a general characterization as
follows. We first create a fiducial object with given values of
redshift ($z_0$), AGN fraction ($\hat{a}_0$) and obscuration
($E(B-V)_0$), from which we produce a set of photometry in all 18
bands of photometry we use. We assume a W2 magnitude $\rm W2_0$ and
convert into upper bounds all the bands where the fiducial flux is
below the corresponding survey limit. Because we want to focus on
systematic uncertainties, we assign the synthetic data points uniform
error bars but we do not actually add any random noise. We then assume
that the photometric redshift estimates have a dispersion of
$0.3~(1+z_0)$ around $z_0$, and proceed to determine $\hat{a}$ in a
grid of redshifts covering the whole interval, determining the median
and the 95.4\% confidence interval of the obtained values. Finally, we
repeat this for different values of $\hat{a}_0$, $E(B-V)_0$, $z_0$ and
W2$_0$. The results are presented in Figures \ref{fg:ahat_acc_sn10}
and \ref{fg:ahat_acc_sn3}. Note that differences between the two
probed photometric depths simply come from the number of bands that
have become upper bounds. In general, these Figures show that for most
parameter combinations, the AGN-host luminosity decomposition is very
stable in the presence of these quite large photometric redshift
errors. A small bias is observed for the high reddening cases
($E(B-V)=5.0$) at all $z_0$ and $W2_0$ values, which is simply caused
by the weak prior on this quantity discussed above, and completely
disappears when we remove it. For W2$_0=15.73$
(Fig. \ref{fg:ahat_acc_sn10}), the $S/N=10$ limit in Bo\"otes, there
is little bias in the median recovered $\hat{a}$ as a function of
$\hat{a}_0$, $E(B-V)_0$ and $z_0$. Error-bars become larger only for
the most galaxy dominated systems (lowest $\hat{a}_0$) at $z_0=2$,
which are exceedingly rare in our sample given our survey depth. For
the fainter case of $\rm W2_0=17.11$ where less bands yield meaningful
constraints, the errors are larger, yet the median of the recovered
$\hat{a}$ values show little bias up to $z_0=1$. At $z_0=2$
significant bias is observed for $\hat{a}\leq 0.6$ and $E(B-V)>0$,
primarily caused by the lack of constraining information in the UV and
optical bands. As mentioned before, however, these systems are
extremely rare in our sample, and hence will not constitute a
significant source of uncertainty in our results.

\section{Additional Spectroscopic Redshifts in the NDWFS Bo\"otes Field}

The four slitmasks that we observed were designed to target
WISE-selected AGN candidates in the Bo\"otes field, though the low
source density of such sources allowed for additional spectroscopic
targets. We filled out the masks with: (1) IRAC-selected AGN
candidates, using the two-color criteria of \citet{stern05} [Column~1
  of table, {\em Target Type} = IRAC AGN]; (2) $z>1$ galaxy cluster
candidates from \citet{eisenhardt08} [{\em Target Type} = IRAC
  cluster]; (3) other 4.5 $\mu$m-selected sources from SDWFS,
typically selected to have [3.6]$-$[4.5] $\geq -0.1$ (AB) which
efficiently selects galaxies at $z > 1.2$ \citep[e.g.,][]{galametz12}
[{\em Target Type} = IRAC]; (4) X-ray sources from XBo\"otes
\citep{murray05,kenter05,brand06} [{\em Target Type} = XBo\"otes]; and
(5) MIPS 24 $\mu$m sources in the field [{\em Target Type} = MIPS].
Given the interest and use of the Bo\"otes field by a broad community,
we include those additional sources here.

Table \ref{tab:additional_redshifts} presents the results for 129
Bo\"otes sources for which we obtained redshifts, including the eleven
targeted sources which are also listed in Table \ref{tab:deimos}.  The
quality flags are defined in \S\ref{ssec:keck}.  Of particular note is
the LRIS mask which confirms cluster 10.220 from the catalog of
\citet{eisenhardt08} to be at $z = 0.96$.

\begin{deluxetable}{l c c c c l}


\tablehead{
  \multicolumn{1}{l}{Selection Criterion} &
  \colhead{$N$} & 
  \colhead{$N~(\hat{a}>0.5)$} & 
  \colhead{Reliability} &
  \colhead{Completeness} &
  \multicolumn{1}{l}{Bands}\\
  \colhead{}&
  \colhead{(deg$^{-2}$)} &
  \colhead{(deg$^{-2}$)} &
  \colhead{Fraction} &
  \colhead{Fraction} &
  \multicolumn{1}{l}{Used}
}

\tablecaption{Surface Density of AGN Candidates at W2$<$17.11\tablenotemark{\dagger}\label{tab:dens_sn3}}
\tabletypesize{\footnotesize}
\tablewidth{0pt}
\tablecolumns{6}

\startdata
\multicolumn{6}{l}{{\textbf{WISE AGN Selection}}}\\
$R_{90}$                          &   \phn130  &    \phn118 & 0.90 & 0.09 & W1, W2\\                        
$R_{75}$                          &   \phn256  &    \phn195 & 0.76 & 0.17 & W1, W2\\                        
$C_{90}$                          &      3702  &       1152 & 0.31 & 0.89 & W1, W2\\                        
$C_{75}$                          &      2117  &    \phn929 & 0.44 & 0.72 & W1, W2\\                        
W1--W2$\ge$0.8\tablenotemark{*}   &      2000  &    \phn901 & 0.45 & 0.70 & W1, W2\\                        
\citet{jarrett11}                 &   \phn469  &    \phn268 & 0.57 & 0.21 & W1, W2, W3\tablenotemark{\ddag}\\  
\citet{mateos12}                  &   \phn391  &    \phn256 & 0.65 & 0.20 & W1, W2, W3\\                    
---.                              & \phn\phn83 & \phn\phn68 & 0.82 & 0.05 & W1, W2, W3, W4\\                
\citet{assef10}                   &      1785  &    \phn841 & 0.47 & 0.65 & W1, W2\\                        
---.                              & \phn\phn44 & \phn\phn43 & 0.97 & 0.03 & W1, W2, W3, W4\\                
\citet{wu12}                      &      3218  &       1109 & 0.34 & 0.86 & W1, W2\\    
\\
\multicolumn{6}{l}{{\textbf{Other Infrared AGN Selection}}}\\
\citet{messias12}                 &    \phn662 &    \phn543 & 0.82 & 0.42 & $K_s$, [4.5], [8.0]\\            
---.                              &    \phn314 &    \phn280 & 0.89 & 0.22 & $K_s$, [4.5], [8.0], MIPS 24$\mu$m\\
\citet{stern05}                   &    \phn986 &    \phn659 & 0.67 & 0.51 & [3.6], [4.5], [5.8], [8.0]\\
\citet{lacy04}                    &      2888  &       1029 & 0.36 & 0.79 & [3.6], [4.5], [5.8], [8.0]\\
\citet{lacy07}                    &      1297  &    \phn735 & 0.57 & 0.57 & [3.6], [4.5], [5.8], [8.0]\\
\enddata

\tablenotetext{\dagger}{The effective W2 limits for the $R_{90}$ and
  $R_{75}$ criteria are 16.26 and 16.45 mag respectively due to the W1
  $S/N>3$ requirement of our sample.}

\tablenotetext{\ddag}{W4 is also used if detected.}

\tablenotetext{*}{This criterion corresponds to the color cut proposed
  by \citetalias{stern12} without the magnitude limit W2$<$15.05.}

\tablecomments{{\footnotesize{WISE AGN selection criteria $R_{90}$,
      $R_{75}$, $C_{90}$ and $C_{75}$ are described by
      eqns. (\ref{eq:rel_sel}) and (\ref{eq:comp_sel}). The remaining
      WISE AGN selection criteria are as follows:
      {\textbf{\cite{jarrett11}}}: W2--W3$>$2.2, W2--W3$>$4.2,
      W1--W2$>$0.1(W2--W3)+0.38, W1--W2$<$1.7 and object is not a
      star; {\textbf{\cite{mateos12} 3-band}}: W2--W3$>$2.157,
      W1--W2$>$0.315(W2--W3)--0.222, W1--W2$<$0.315(W2--W3)+0.796;
      {\textbf{\cite{mateos12} 4-band}}: W3--W4$\geq$1.76,
      W1--W2$>$0.50(W3--W4)--0.405, W1--W2$<$0.50(W3--W4)+0.979;
      {\textbf{\cite{assef10} 2-band}}: W1--W2$>$0.85;
      {\textbf{\cite{assef10} 4-band}}: W3--W4$>$2.1, W1--W2$>$0.85,
      W1--W2$>$1.67(W3--W4)--3.41; {\textbf{\cite{wu12}}}:
      W1--W2$>$0.57}}. For the other infrared AGN selection criteria
  we refer the reader to the original studies.}

\end{deluxetable}

\begin{deluxetable}{l c c c c l}


\tablehead{
  \multicolumn{1}{l}{Selection Criterion} &
  \colhead{$N$} & 
  \colhead{$N~(\hat{a}>0.5)$} & 
  \colhead{Reliability} &
  \colhead{Completeness} &
  \multicolumn{1}{l}{Bands}\\
  \colhead{}&
  \colhead{(deg$^{-2}$)} &
  \colhead{(deg$^{-2}$)} &
  \colhead{Fraction} &
  \colhead{Fraction} &
  \multicolumn{1}{l}{Used}
}

\tablecaption{Surface Density of AGN Candidates at W2$<$15.73\label{tab:dens_sn10}}
\tabletypesize{\footnotesize}
\tablewidth{0pt}
\tablecolumns{6}

\startdata
\multicolumn{6}{l}{{\textbf{WISE AGN Selection}}}\\
$R_{90}$                          &  \phn117  &    106 & 0.90 & 0.53 & W1, W2\\                        
$R_{75}$                          &  \phn176  &    133 & 0.76 & 0.67 & W1, W2\\                        
$C_{90}$                          &  \phn439  &    177 & 0.40 & 0.88 & W1, W2\\                        
$C_{75}$                          &  \phn194  &    144 & 0.74 & 0.72 & W1, W2\\                        
 W1--W2$\ge$0.8\tablenotemark{*}  &  \phn182  &    139 & 0.77 & 0.69 & W1, W2\\                        
\citet{jarrett11}                 &  \phn166  &    128 & 0.77 & 0.64 & W1, W2, W3\tablenotemark{\ddag}\\  
\citet{mateos12}                  &  \phn161  &    129 & 0.80 & 0.64 & W1, W2, W3\\                    
---.                              &\phn\phn69 & \phn56 & 0.80 & 0.28 & W1, W2, W3, W4\\                
\citet{assef10}                   &  \phn161  &    130 & 0.81 & 0.65 & W1, W2\\                        
---.                              &\phn\phn39 & \phn38 & 0.98 & 0.19 & W1, W2, W3, W4\\                
\citet{wu12}                      &  \phn347  &    170 & 0.49 & 0.85 & W1, W2\\   
\\
\multicolumn{6}{l}{{\textbf{Other Infrared AGN Selection}}}\\
\citet{messias12}                 &  \phn166  &    152 & 0.91 & 0.76 & $K_s$, [4.5], [8.0]\\                
---.                              &  \phn107  &    102 & 0.96 & 0.51 & $K_s$, [4.5], [8.0], MIPS 24$\mu$m\\ 
\citet{stern05}                   &  \phn194  &    157 & 0.81 & 0.79 & [3.6], [4.5], [5.8], [8.0]\\      
\citet{lacy04}                    &  \phn484  &    183 & 0.38 & 0.92 & [3.6], [4.5], [5.8], [8.0]\\      
\citet{lacy07}                    &  \phn262  &    172 & 0.66 & 0.86 & [3.6], [4.5], [5.8], [8.0]\\      
\enddata

\tablenotetext{\ddag}{W4 is also used if detected.}

\tablenotetext{*}{This criterion corresponds to the color cut proposed
  by \citetalias{stern12} without the magnitude limit W2$<$15.05.}

\end{deluxetable}

\begin{deluxetable}{l c c c c l}


\tablehead{
  \multicolumn{1}{l}{Selection Criterion} &
  \colhead{$N$} & 
  \colhead{$N~(\hat{a}>0.5)$} & 
  \colhead{Reliability} &
  \colhead{Completeness} &
  \multicolumn{1}{l}{Bands}\\
  \colhead{}&
  \colhead{(deg$^{-2}$)} &
  \colhead{(deg$^{-2}$)} &
  \colhead{Fraction} &
  \colhead{Fraction} &
  \multicolumn{1}{l}{Used}
}

\tablecaption{Surface Density of AGN Candidates at W2$<$15.05\label{tab:dens_cosmos}}
\tabletypesize{\footnotesize}
\tablewidth{0pt}
\tablecolumns{6}

\startdata
\multicolumn{6}{l}{{\textbf{WISE AGN Selection}}}\\
$R_{90}$               &  66 & 59 & 0.90 & 0.77 & W1, W2\\                        
$R_{75}$               &  87 & 65 & 0.75 & 0.84 & W1, W2\\                        
$C_{90}$               & 120 & 69 & 0.57 & 0.89 & W1, W2\\                        
$C_{75}$               &  64 & 59 & 0.93 & 0.77 & W1, W2\\                        
\citetalias{stern12}   &  62 & 58 & 0.94 & 0.75 & W1, W2\\                        
\citet{jarrett11}      &  66 & 59 & 0.90 & 0.77 & W1, W2, W3\tablenotemark{\ddag}\\  
\citet{mateos12}       &  65 & 60 & 0.92 & 0.78 & W1, W2, W3\\                    
---.                   &  48 & 39 & 0.82 & 0.51 & W1, W2, W3, W4\\                
\citet{assef10}        &  57 & 55 & 0.96 & 0.71 & W1, W2\\                        
---.                   &  29 & 29 & 1.00 & 0.37 & W1, W2, W3, W4\\                
\citet{wu12}           &  99 & 67 & 0.68 & 0.87 & W1, W2\\                          
\\
\multicolumn{6}{l}{{\textbf{Other Infrared AGN Selection}}}\\
\citet{messias12}      &  67 & 63 & 0.94 & 0.82 & $K_s$, [4.5], [8.0]\\                
---.                   &  48 & 47 & 0.97 & 0.60 & $K_s$, [4.5], [8.0], MIPS 24$\mu$m\\ 
\citet{stern05}        &  74 & 66 & 0.89 & 0.85 & [3.6], [4.5], [5.8], [8.0]\\      
\citet{lacy04}         & 123 & 70 & 0.56 & 0.90 & [3.6], [4.5], [5.8], [8.0]\\      
\citet{lacy07}         &  91 & 68 & 0.75 & 0.89 & [3.6], [4.5], [5.8], [8.0]\\      
\enddata

\tablenotetext{\ddag}{W4 is also used if detected.}

\end{deluxetable}

\begin{deluxetable}{l c c c c c c}

\tablehead{
  \multicolumn{1}{l}{Sample} & 
  \colhead{$\Delta z$} & 
  \colhead{$\Delta z_{95}$} &
  \colhead{bias$/(1+z)$} & 
  \colhead{bias$^{95\%}/(1+z)$} &
  \colhead{N$_{\rm AGN}$} &
  \colhead{$z_{s}$ Fraction}
}

\tablecaption{Photometric Redshifts\label{tab:photoz}}
\tabletypesize{\small}
\tablewidth{0pt}
\tablecolumns{7}

\startdata

\multicolumn{7}{l}{{\textbf{Full W2 Depth}}}\\
$R_{90}$ &   0.27 &   0.20 &   0.15 &   0.14 &    618 &     0.56\\
$R_{75}$ &   0.29 &   0.21 &   0.14 &   0.13 &    839 &     0.44\\
$C_{90}$ &   0.31 &   0.23 &   0.13 &   0.12 &   1668 &     0.13\\
$C_{75}$ &   0.29 &   0.23 &   0.17 &   0.16 &   1360 &     0.10\\
\multicolumn{7}{l}{{\textbf{W2$<$15.73}}}\\
$R_{90}$ &   0.27 &   0.20 &   0.14 &   0.13 &    595 &     0.61\\
$R_{75}$ &   0.29 &   0.20 &   0.13 &   0.12 &    731 &     0.57\\
$C_{90}$ &   0.29 &   0.19 &   0.09 &   0.09 &    890 &     0.48\\
$C_{75}$ &   0.26 &   0.20 &   0.13 &   0.12 &    740 &     0.48\\
\multicolumn{7}{l}{{\textbf{W2$<$15.73 and $I$$<$20}}}\\
$R_{90}$ &   0.27 &   0.20 &   0.13 &   0.12 &    420 &     0.85\\
$R_{75}$ &   0.30 &   0.20 &   0.12 &   0.11 &    507 &     0.84\\
$C_{90}$ &   0.30 &   0.19 &   0.09 &   0.09 &    597 &     0.75\\
$C_{75}$ &   0.27 &   0.20 &   0.13 &   0.12 &    477 &     0.82\\
\multicolumn{7}{l}{{\textbf{W2$<$15.73 and $I$$<$20 and $E(B-V)$$>$0.5}}}\\
$R_{90}$ &   0.25 &   0.11 &   0.10 &   0.11 &     68 &     0.77\\
$R_{75}$ &   0.42 &   0.12 &   0.09 &   0.10 &     90 &     0.84\\
$C_{90}$ &   0.42 &   0.11 &   0.04 &   0.05 &    134 &     0.76\\
$C_{75}$ &   0.24 &   0.12 &   0.11 &   0.12 &     78 &     0.78\\
\enddata

\tablecomments{The table shows the measured photometric redshift
  dispersions $\Delta z$ and $\Delta z_{95}$ (see
  \S\ref{ssec:photoz_acc} for details), as well as the mean bias of
  each sample, measured as $<|z_{\rm phot}-z_{\rm spec}|>$ for all
  objects and limited to the 95\% with the best photometric redshift
  determination. Note that the numbers only reflect the statistics for
  objects with $\hat{a}>0.5$ to avoid improved accuracies due to
  contamination by inactive galaxies.}

\end{deluxetable}

\pagestyle{empty} 

\begin{deluxetable}{l c c c c c c c c c c c}

\rotate

\tablehead{
  \colhead{Name} & 
  \colhead{R.A.} & 
  \colhead{Dec.} & 
  \colhead{$I$ (mag)} &
  \colhead{W2 (mag)} &
  \colhead{W1--W2} &
  \colhead{$\hat{a}$} &
  \colhead{$E(B-V)$} &
  \colhead{$z$} &
  \colhead{Q} &
  \colhead{Selection} &
  \colhead{Notes}
}

\tablecaption{Summary of Obscured AGN Candidates Observed with Keck/DEIMOS\label{tab:deimos}}
\tabletypesize{\footnotesize}
\tablewidth{0pt}
\tablecolumns{12}

\startdata
W1427+3400 &14:27:54.57 &34:00:43.31 &21.90 &15.81 &0.82 &0.386 $\pm$ 0.091 & 0.41$\pm$ 0.11 &  1.293  & A & $C_{75}$   & [O\,{\sc ii}]\\       
W1427+3403 &14:27:47.16 &34:03:41.84 &21.24 &15.76 &0.50 &0.209 $\pm$ 0.117 & 0.33$\pm$ 0.12 &  1.137  & A & {\it{None}}& CaHK\\
W1427+3408 &14:27:17.93 &34:08:28.60 &21.71 &14.99 &2.08 &0.989 $\pm$ 0.009 & 0.88$\pm$ 0.04 &  1.158  & A & $R_{90}$   & [O\,{\sc ii}]\\       
W1428+3359 &14:28:12.31 &33:59:25.13 &23.22 &15.15 &1.02 &0.685 $\pm$ 0.027 & 0.87$\pm$ 0.05 &  1.343  & B & $R_{90}$   & [O\,{\sc ii}]\\       
W1429+3529 &14:29:54.83 &35:29:04.08 &21.97 &15.77 &0.71 &0.455 $\pm$ 0.082 & 0.20$\pm$ 0.08 & (1.3)   & F & $C_{90}$   & \\       
W1430+3525 &14:30:31.69 &35:25:17.78 &20.64 &15.07 &1.54 &0.944 $\pm$ 0.012 & 0.45$\pm$ 0.06 &  1.106  & A & $R_{90}$   & [O\,{\sc ii}],CaHK\\       
W1430+3530 &14:30:00.50 &35:30:55.01 &19.42 &15.55 &1.06 &0.497 $\pm$ 0.044 & 0.25$\pm$ 0.07 & (1.3)   & F & $R_{75}$   & \\       
W1431+3525 &14:31:06.26 &35:25:46.24 &23.63 &15.60 &1.88 &1.000 $\pm$ 0.004 & 0.78$\pm$ 0.05 & (1.3)   & F & $R_{90}$   & [O\,{\sc ii}]\\       
W1431+3528 &14:31:31.38 &35:28:38.21 &23.62 &15.47 &1.77 &0.990 $\pm$ 0.004 & 0.89$\pm$ 0.04 &  1.343  & A & $R_{90}$   & [O\,{\sc ii}]\\       
W1432+3523 &14:32:23.02 &35:23:21.41 &19.03 &14.33 &1.01 &0.742 $\pm$ 0.006 & 0.50$\pm$ 0.03 &  0.258  & A & $R_{90}$   & CaHK,H$\alpha$,[N\,{\sc ii}]\\       
W1432+3525 &14:32:37.30 &35:25:12.56 &21.94 &15.54 &0.72 &0.583 $\pm$ 0.036 & 0.75$\pm$ 0.05 &  1.117  & A & $C_{90}$   & Mg\,{\sc ii} absn,[O\,{\sc ii}],D4000\\    
W1432+3526 &14:32:22.61 &35:26:46.88 &22.85 &15.01 &1.53 &0.954 $\pm$ 0.006 & 0.79$\pm$ 0.05 &  1.436  & B & $R_{90}$   & [O\,{\sc ii}]\\       
\enddata

\tablecomments{The AGN selection criteria listed in the last column is
  the least inclusive one met. For sources that failed to yield
  spectroscopic redshifts in these observations, we list the
  photometric redshift in parenthesis. Coordinates are J2000.}

\end{deluxetable}

\begin{deluxetable}{l c c c c c l}

\rotate

\tablehead{
   \multicolumn{1}{l}{Target Type} &
   \colhead{R.A.} &
   \colhead{Dec.} &
   \colhead{$z$} &
   \colhead{Q} &
   \colhead{Slitmask(s)} &
   \multicolumn{1}{l}{Notes}
}

\tablecaption{Additional Results from Keck Observations.\label{tab:additional_redshifts}}
\tabletypesize{\small}
\tablewidth{0pt}
\tablecolumns{7}

\startdata
IRAC		        & 14:27:13.39 & +34:09:04.5 & 1.343 & A & C[36] & Mg\,{\sc ii} absorption,[O\,{\sc ii}] \\
IRAC AGN              	& 14:27:14.32 & +34:09:01.3 & 1.692 & B & C[14] & QSO: Mg\,{\sc ii} (w/ Mg\,{\sc ii} absorption system at z=1.342) \\
MIPS		        & 14:27:14.63 & +34:08:46.6 & 1.343 & B & C[33] & QSO: Mg\,{\sc ii} \\
MIPS		        & 14:27:14.77 & +34:06:09.0 & 1.151 & A & C[28] & Mg\,{\sc ii} absorption,[O\,{\sc ii}],D4000 \\
MIPS		        & 14:27:17.40 & +34:07:26.9 & 1.082 & A & C[31] & [O\,{\sc ii}],[Ne\,{\sc iii}] \\
WISE AGN	        & 14:27:17.92 & +34:08:28.6 & 1.158 & A & C[03] & [O\,{\sc ii}] \\
MIPS		        & 14:27:18.37 & +34:09:01.3 & 0.130 & A & C[34] & [O\,{\sc iii}],H$\alpha$ \\
IRAC		        & 14:27:19.18 & +34:06:47.0 & 1.213 & B & C[37] & [O\,{\sc ii}],D4000 \\
IRAC		        & 14:27:20.76 & +34:06:42.8 & 0.305 & A & C[38] & H$\beta$,[O\,{\sc iii}],H$\alpha$ \\
MIPS		        & 14:27:23.95 & +34:06:15.5 & 0.190 & A & C[29] & H$\beta$,[O\,{\sc iii}],H$\alpha$ \\
XBo\"otes		& 14:27:24.80 & +34:05:12.7 & 1.753 & A & C[15] & QSO: C\,{\sc iii}],Mg\,{\sc ii} \\
IRAC		        & 14:27:27.55 & +34:05:38.1 & 1.483 & A & C[40] & B2900,[O\,{\sc ii}] \\
MIPS		        & 14:27:27.56 & +34:04:13.7 & 0.126 & A & C[23] & H$\beta$,[O\,{\sc iii}],H$\alpha$ \\
IRAC		        & 14:27:27.90 & +34:06:25.5 & 1.690 & B & C[41] & Mg\,{\sc ii} absorption,[O\,{\sc ii}] (do not see past 1$\mu$m in Allslits, to confirm [O\,{\sc ii}]) \\
MIPS		        & 14:27:29.76 & +34:04:34.8 & 1.065 & A & C[25] & [O\,{\sc ii}],D4000 \\
MIPS		        & 14:27:30.15 & +34:06:05.0 & 0.191 & A & C[27] & H$\beta$,[O\,{\sc iii}],H$\alpha$ \\
IRAC AGN              	& 14:27:31.50 & +34:03:28.6 & 2.192 & A & C[13] & QSO2: C\,{\sc iv},He\,{\sc ii},C\,{\sc iii}] \\
IRAC		        & 14:27:32.07 & +34:03:19.4 & 1.413 & A & C[42] & [O\,{\sc ii}] \\
MIPS		        & 14:27:32.42 & +34:08:08.0 & 0.872 & A & C[32] & [O\,{\sc ii}],D4000 \\
IRAC		        & 14:27:41.87 & +34:04:55.0 & 1.524 & A & C[43] & Mg\,{\sc ii} absorption,[O\,{\sc ii}] \\
IRAC AGN              	& 14:27:43.36 & +34:03:06.1 & 3.134 & A & C[12] & QSO: Ly$\alpha$,C\,{\sc iv},C\,{\sc iii}] (self-absorbed) \\
WISE AGN        	& 14:27:47.17 & +34:03:41.2 & 1.137 & A & C[02] & CaHK \\
serendip              	& 14:27:47.26 & +34:03:40.8 & 1.623 & B & C[02] & [O\,{\sc ii}] - v. ft. serendipitous \\
IRAC AGN              	& 14:27:47.98 & +34:02:44.1 & 3.276 & A & C[11] & QSO: Ly$\alpha$,C\,{\sc iv},C\,{\sc iii}] \\
IRAC		        & 14:27:48.28 & +34:03:11.7 & 1.427 & A & C[45] & [O\,{\sc ii}] \\
WISE AGN        	& 14:27:54.60 & +34:00:43.0 & 1.293 & A & C[01] & [O\,{\sc ii}] \\
MIPS		        & 14:27:55.18 & +34:02:41.9 & 0.643 & A & C[22] & [O\,{\sc ii}],H$\beta$,[O\,{\sc iii}] \\
IRAC		        & 14:27:57.49 & +34:03:50.8 & 0.494 & A & C[35] & [O\,{\sc ii}],CaHK,H$\beta$,H$\alpha$ \\
IRAC		        & 14:27:59.64 & +34:03:07.3 & 1.621 & A & C[49] & [O\,{\sc ii}] \\
MIPS		        & 14:28:03.55 & +33:59:23.4 & 0.756 & A & C[18] & [O\,{\sc ii}],CaHK,H$\beta$ \\
MIPS		        & 14:28:03.65 & +34:02:25.3 & 0.679 & A & C[20] & [O\,{\sc ii}],H$\beta$,[O\,{\sc iii}] \\
MIPS		        & 14:28:04.84 & +34:02:27.8 & 0.458 & A & C[21] & [O\,{\sc ii}],H$\beta$,[O\,{\sc iii}],H$\alpha$ \\
MIPS		        & 14:28:07.73 & +33:58:28.9 & 1.262 & A & C[17] & QSO: Mg\,{\sc ii},[O\,{\sc ii}],[Ne\,{\sc iii}] \\
MIPS		        & 14:28:09.59 & +34:01:31.6 & 0.627 & A & C[19] & [O\,{\sc ii}],H$\beta$,[O\,{\sc iii}] \\
MIPS		        & 14:28:09.71 & +33:58:06.6 & 1.246 & A & C[16] & [O\,{\sc ii}] \\
XBo\"otes	        & 14:28:10.07 & +34:28:19.4 & 1.234 & A & F     & AGN: C\,{\sc iii}],[O\,{\sc ii}] \\
IRAC		        & 14:28:11.28 & +34:00:54.6 & 1.484 & A & C[53] & [O\,{\sc ii}] \\
WISE AGN        	& 14:28:12.30 & +33:59:25.6 & 1.343 & B & C[00] & [O\,{\sc ii}] \\
IRAC	                & 14:28:14.26 & +34:29:22.5 & 0.607 & A & F     & AGN \\
IRAC cluster            & 14:28:14.91 & +34:28:28.2 & 1.08  & B & F     & D4000 \\
IRAC cluster            & 14:28:15.24 & +34:27:35.1 & 1.023 & B & F     & [O\,{\sc ii}] \\
IRAC	                & 14:28:19.56 & +34:23:10.0 & 0.363 & A & F     & [O\,{\sc ii}],H$\alpha$ \\
IRAC cluster            & 14:28:21.26 & +34:28:52.3 & 0.962 & A & F     & Mg\,{\sc ii} absorption,[O\,{\sc ii}],D4000,H$\delta$ \\
IRAC cluster            & 14:28:22.15 & +34:27:36.4 & 1.140 & A & F     & [O\,{\sc ii}],CaHK \\
IRAC cluster            & 14:28:22.34 & +34:27:21.6 & 0.963 & A & F     & CaHK \\
IRAC cluster            & 14:28:23.63 & +34:27:17.5 & 0.96  & B & F     & D4000 \\
WISE AGN            	& 14:28:25.03 & +34:22:37.2 & 1.518 & B & F     & [O\,{\sc ii}] \\
IRAC cluster            & 14:28:25.06 & +34:26:59.5 & 0.963 & B & F     & CaK,D4000 \\
IRAC cluster            & 14:28:25.86 & +34:25:33.4 & 0.95  & B & F     & D4000 \\
XBo\"otes	        & 14:28:29.06 & +34:25:45.6 & 1.036 & A & F     & AGN: broad Mg\,{\sc ii},[Ne\,{\sc v}],[O\,{\sc ii}],[O\,{\sc iii}] \\
IRAC	                & 14:28:30.21 & +34:27:33.8 & 0.424 & A & F     & Mg\,{\sc ii} absorption,H$\alpha$ \\
IRAC	                & 14:28:31.10 & +34:24:28.7 & 3.333 & B & F     & asymmetric Ly$\alpha$, Ly$\alpha$ break \\
IRAC	                & 14:28:37.32 & +34:24:50.5 & 1.049 & A & F     & [O\,{\sc ii}],D4000 \\
IRAC	                & 14:28:40.39 & +34:24:21.0 & 0.694 & A & F     & AGN: broad Mg\,{\sc ii},narrow H$\beta$,[O\,{\sc iii}] \\
MIPS      		& 14:29:55.16 & +35:30:26.2 & 0.632 & A & A[45] & [O\,{\sc ii}],H$\beta$,[O\,{\sc iii}] (H$\beta$ w/ 2nd component) \\
XBo\"otes      		& 14:29:57.10 & +35:30:54.5 & 0.264 & A & A[15] & QSO: H$\beta$,[O\,{\sc iii}],H$\alpha$ \\
serendip              	& 14:29:57.69 & +35:27:42.4 & 0.492 & A & A[39] & [O\,{\sc ii}],[O\,{\sc iii}] \\      
MIPS      		& 14:29:57.94 & +35:27:42.1 & 0.547 & A & A[39] & [O\,{\sc ii}],H$\beta$,[O\,{\sc iii}] \\
IRAC		        & 14:29:58.94 & +35:29:27.9 & 1.576 & B & A[51] & [O\,{\sc ii}] \\
IRAC		        & 14:30:06.07 & +35:31:10.6 & 1.275 & B & A[52] & CaHK \\
MIPS      		& 14:30:07.83 & +35:29:28.0 & 1.004 & A & A[43] & [O\,{\sc ii}],D4000 \\
IRAC		        & 14:30:08.78 & +35:30:51.3 & 1.092 & A & A[53] & [O\,{\sc ii}],D4000 \\
MIPS      		& 14:30:09.20 & +35:28:54.3 & 0.059 & A & A[41] & H$\beta$,[O\,{\sc iii}],H$\alpha$,[S\,{\sc iii}] \\
IRAC AGN              	& 14:30:09.55 & +35:28:22.3 & 0.978 & A & A[14] & QSO-2: lots of lines, including [Ne\,{\sc v}]3426 \\
IRAC		        & 14:30:11.03 & +35:27:08.1 & 3.156 & B & A[54] & Ly$\alpha$ (serendipitous?) \\
MIPS      		& 14:30:12.50 & +35:27:14.6 & 0.177 & A & A[36] & H$\beta$,[O\,{\sc iii}],H$\alpha$ \\
MIPS      		& 14:30:13.14 & +35:26:53.7 & 0.362 & A & A[33] & CaHK \\
MIPS      		& 14:30:14.33 & +35:26:49.4 & 1.481 & A & A[32] & AGN: [Ne\,{\sc v}],[O\,{\sc ii}],[Ne\,{\sc iii}] \\
MIPS      		& 14:30:16.54 & +35:30:41.9 & 0.792 & A & A[47] & [O\,{\sc ii}],H$\beta$,[O\,{\sc iii}] \\
IRAC AGN              	& 14:30:19.78 & +35:27:04.7 & 1.268 & A & A[13] & AGN: [Ne\,{\sc v}]3426,[O\,{\sc ii}] \\
MIPS      		& 14:30:20.75 & +35:26:50.3 & 0.681 & A & A[31] & CaHK \\
MIPS      		& 14:30:21.18 & +35:29:12.7 & 0.0   & A & A[42] & M-star \\
IRAC		        & 14:30:25.05 & +35:27:28.8 & 0.351 & A & A[49] & CaHK,H$\alpha$,[N\,{\sc ii}] (AGN based on [N\,{\sc ii}]/H$\alpha$ ratio?) \\
serendip             	& 14:30:25.56 & +35:26:28.4 & 0.664 & A & A[29] & [O\,{\sc ii}],H$\beta$,[O\,{\sc iii}] \\
MIPS      		& 14:30:26.03 & +35:28:42.4 & 0.264 & A & A[40] & H$\beta$,[O\,{\sc iii}],H$\alpha$ ([O\,{\sc ii}] in 2nd order?) \\
MIPS      		& 14:30:26.34 & +35:26:22.2 & 0.940 & A & A[29] & [O\,{\sc ii}],CaHK \\
MIPS      		& 14:30:26.96 & +35:26:05.8 & 0.349 & A & A[27] & [O\,{\sc ii}],CaHK,H$\beta$,[O\,{\sc iii}],H$\alpha$ \\
MIPS      		& 14:30:30.74 & +35:25:42.8 & 1.024 & A & A[24] & [O\,{\sc ii}] \\
WISE AGN 	      	& 14:30:31.71 & +35:25:17.8 & 1.106 & A & A[00] & [O\,{\sc ii}],CaHK \\ 
serendip              	& 14:30:31.92 & +35:25:17.4 & 0.153 & A & A[00] & [O\,{\sc iii}],H$\alpha$,[S\,{\sc ii}] \\
MIPS      		& 14:30:32.32 & +35:27:06.2 & 1.025 & A & A[35] & [O\,{\sc ii}],D4000 \\
MIPS      		& 14:30:37.25 & +35:26:37.1 & 1.012 & A & A[30] & QSO: broad Mg\,{\sc ii},[O\,{\sc ii}],H$\beta$,[O\,{\sc iii}] \\
serendip             	& 14:30:37.26 & +35:27:38.2 & 0.524 & A & A[38] & [O\,{\sc ii}],H$\beta$,[O\,{\sc iii}] (or target?) \\
IRAC		        & 14:30:38.10 & +35:24:47.1 & 1.305 & A & A[56] & [O\,{\sc ii}] \\
WISE AGN               	& 14:30:42.92 & +35:28:14.7 & 1.395 & B & A[02] & [O\,{\sc ii}] \\
MIPS      		& 14:30:46.02 & +35:24:50.4 & 0.308 & A & A[20] & [O\,{\sc ii}],H$\beta$,[O\,{\sc iii}],H$\alpha$ \\
MIPS      		& 14:30:48.46 & +35:27:02.4 & 0.375 & A & A[34] & CaHK,H$\alpha$ \\
IRAC AGN              	& 14:30:51.09 & +35:24:08.0 & 2.043 & A & A[11] & QSO: C\,{\sc iv},C\,{\sc iii}],Mg\,{\sc ii} \\
serendip              	& 14:30:51.34 & +35:24:07.2 & 0.0   & A & A[11] & star: H$\alpha$,CaT absorption \\
IRAC AGN              	& 14:30:51.46 & +35:25:49.4 & 0.321 & A & A[12] & CaHK \\
MIPS      		& 14:30:56.51 & +35:25:40.8 & 0.720 & A & A[23] & [O\,{\sc ii}],Balmer absorption,H$\beta$,[O\,{\sc iii}] \\
MIPS      		& 14:31:00.30 & +35:25:46.0 & 0.084 & A & A[25] & H$\beta$,[O\,{\sc iii}],H$\alpha$ \\
MIPS      		& 14:31:01.91 & +35:24:48.6 & 0.650 & A & A[19] & AGN: [Ne\,{\sc v}]3426,[O\,{\sc ii}],H$\beta$,[O\,{\sc iii}] \\
MIPS      		& 14:31:02.47 & +35:24:34.5 & 0.651 & A & A[18] & [O\,{\sc ii}],H$\beta$,[O\,{\sc iii}] \\
serendip              	& 14:31:05.86 & +35:24:00.6 & 0.414 & A & A[48] & H$\beta$,[O\,{\sc iii}],H$\alpha$ \\
IRAC		        & 14:31:06.00 & +35:23:59.4 & 0.429 & A & A[48] & H$\beta$,[O\,{\sc iii}] \\
WISE AGN 		& 14:31:31.39 & +35:28:38.3 & 1.343 & A & B[03] & [O\,{\sc ii}] \\
IRAC AGN              	& 14:31:34.52 & +35:29:23.3 & 1.210 & A & B[15] & QSO-2: lots of lines \\
MIPS		        & 14:31:34.83 & +35:27:46.6 & 0.485 & A & B[33] & [O\,{\sc ii}],H$\beta$ \\
serendip             	& 14:31:36.26 & +35:29:18.9 & 0.163 & A & B[40] & H$\beta$,[O\,{\sc iii}],H$\alpha$ \\
IRAC		        & 14:31:39.75 & +35:28:01.4 & 1.028 & A & B[42] & [O\,{\sc ii}] \\
MIPS		        & 14:31:41.00 & +35:28:11.9 & 0.958 & A & B[35] & [O\,{\sc ii}] \\
MIPS		        & 14:31:41.18 & +35:29:44.8 & 0.163 & A & B[41] & [O\,{\sc iii}],[N\,{\sc ii}],[S\,{\sc ii}] \\
serendip              	& 14:31:41.32 & +35:28:09.6 & 1.101 & A & B[35] & [O\,{\sc ii}] \\
MIPS		        & 14:31:41.87 & +35:27:49.5 & 0.820 & A & B[34] & [O\,{\sc ii}],H$\beta$,[O\,{\sc iii}] \\
MIPS		        & 14:31:44.23 & +35:27:43.3 & 0.0   & A & B[32] & star: CaT \\
MIPS		        & 14:31:47.43 & +35:28:51.2 & 0.243 & A & B[39] & [O\,{\sc ii}],H$\alpha$ \\
MIPS		        & 14:31:48.57 & +35:28:15.5 & 1.215 & A & B[36] & [O\,{\sc ii}] \\
IRAC		        & 14:31:52.06 & +35:28:21.2 & 0.765 & A & B[43] & [O\,{\sc ii}],CaHK,[O\,{\sc iii}] \\
serendip             	& 14:31:53.61 & +35:26:26.5 & 0.0   & A & B[49] & M-star \\
IRAC AGN              	& 14:31:55.50 & +35:26:40.3 & 0.911 & A & B[14] & QSO: [O\,{\sc ii}],CaHK,[O\,{\sc iii}],broad H$\alpha$ \\
serendip              	& 14:31:58.55 & +35:25:51.6 & 1.136 & A & B[51] & [O\,{\sc ii}] \\
IRAC		        & 14:31:58.85 & +35:25:49.0 & 2.957 & A & B[51] & ft. QSO: Ly$\alpha$,C\,{\sc iv},He\,{\sc ii},C\,{\sc iii}] \\
MIPS		        & 14:31:59.27 & +35:26:46.2 & 0.369 & A & B[26] & [O\,{\sc ii}],H$\alpha$ \\
IRAC		        & 14:31:59.54 & +35:28:21.9 & 0.557 & A & B[44] & CaHK \\
MIPS		        & 14:32:01.73 & +35:28:46.0 & 0.392 & A & B[38] & H$\beta$,[O\,{\sc iii}],H$\alpha$ \\
MIPS		        & 14:32:03.08 & +35:26:46.7 & 0.336 & A & B[27] & H$\beta$,[O\,{\sc iii}],H$\alpha$ \\
serendip             	& 14:32:04.97 & +35:26:42.4 & 0.438 & A & B[27] & [O\,{\sc ii}],CaHK,H$\alpha$ \\
MIPS		        & 14:32:06.77 & +35:25:08.0 & 0.732 & A & B[24] & [O\,{\sc ii}],H$\beta$,[O\,{\sc iii}] \\
IRAC AGN              	& 14:32:11.50 & +35:25:34.6 & 0.307 & A & B[12] & CaHK,H$\beta$,[O\,{\sc iii}],H$\alpha$ \\
IRAC		        & 14:32:13.37 & +35:27:10.5 & 0.0   & A & B[45] & star: CaT \\
IRAC		        & 14:32:17.71 & +35:25:21.7 & 0.368 & A & B[46] & [O\,{\sc ii}],CaHK,H$\beta$,H$\alpha$ \\
serendip             	& 14:32:18.14 & +35:25:19.3 & 1.136 & A & B[46] & [O\,{\sc ii}] \\
MIPS		        & 14:32:19.72 & +35:24:26.5 & 0.666 & A & B[23] & [O\,{\sc ii}],H$\beta$,[O\,{\sc iii}] \\
WISE AGN 		& 14:32:22.64 & +35:26:46.7 & 1.436 & B & B[02] & [O\,{\sc ii}] (bad slitlet) \\
WISE AGN 		& 14:32:23.06 & +35:23:21.4 & 0.258 & A & B[00] & CaHK,H$\alpha$,[N\,{\sc ii}] (likely AGN from H$\alpha$/[N\,{\sc ii}] ratio) \\
MIPS		        & 14:32:26.08 & +35:27:14.7 & 0.662 & A & B[28] & [O\,{\sc ii}],H$\beta$ \\
MIPS		        & 14:32:27.84 & +35:22:42.0 & 0.469 & A & B[20] & [O\,{\sc ii}],H$\beta$,[O\,{\sc iii}],H$\alpha$ \\
MIPS		        & 14:32:28.30 & +35:27:35.1 & 0.380 & A & B[30] & CaHK,H$\beta$,[O\,{\sc iii}],H$\alpha$ \\
MIPS		        & 14:32:29.30 & +35:22:36.0 & 1.011 & A & B[18] & [O\,{\sc ii}],D4000 \\
IRAC		        & 14:32:29.80 & +35:26:29.5 & 1.136 & A & B[48] & Mg\,{\sc ii} absorption,[O\,{\sc ii}] \\
XBo\"otes       	& 14:32:32.03 & +35:26:26.6 & 1.077 & A & B[16] & [O\,{\sc ii}],CaHK \\
IRAC AGN              	& 14:32:36.49 & +35:25:39.4 & 1.070 & A & B[13] & QSO: broad Mg\,{\sc ii},[O\,{\sc ii}],broad Balmer lines \\
WISE AGN 		& 14:32:37.33 & +35:25:12.6 & 1.117 & A & B[01] & Mg\,{\sc ii} absorption,[O\,{\sc ii}],D4000 \\
MIPS		        & 14:32:38.38 & +35:21:54.2 & 0.495 & A & B[17] & [O\,{\sc ii}],H$\beta$,[O\,{\sc iii}] \\
IRAC AGN              	& 14:32:39.41 & +35:23:45.0 & 1.979 & B & B[11] & QSO: Mg\,{\sc ii} \\
serendip              	& 14:32:39.81 & +35:21:55.6 & 0.835 & B & B[10] & [O\,{\sc ii}] \\
IRAC AGN              	& 14:32:40.04 & +35:21:54.5 & 0.543 & A & B[10] & BL AGN: lots of lines \\
MIPS		        & 14:32:41.29 & +35:26:04.5 & 0.071 & A & B[25] & H$\beta$,[O\,{\sc iii}],H$\alpha$ \\
MIPS		        & 14:32:45.09 & +35:23:23.4 & 0.601 & A & B[22] & [O\,{\sc ii}],H$\beta$ \\
\enddata

\tablecomments{Spectroscopic measurements for 129 additional
  sources. Coordinates shown are in J2000. Q indicates the quality of
  the redshift (see \S\ref{ssec:keck} for details). Masks A, B and C
  were observed with DEIMOS, and the bracketed numbers indicate the
  DEIMOS slitlet number. Mask F was observed with LRIS.}

\end{deluxetable}

\end{document}